\documentclass[prd,aps,preprint,showpacs,nofootinbib]{revtex4}
\usepackage{epsfig}
\usepackage{amsmath}
\usepackage{color}
\usepackage{latexsym}
\usepackage{amssymb}
\usepackage{dsfont}
\usepackage{multirow}

\newcommand{\bea}{\begin{eqnarray}}
\newcommand{\eea}{\end{eqnarray}}
\newcommand{\be}{\begin{equation}}
\newcommand{\ee}{\end{equation}}

\newcommand{\ud}{\mathrm{d}}

\newlength\savedwidth

\newcommand{\uvec}[1]{\boldsymbol{#1}}

\begin{document}
\preprint{}

\title{The Quark Orbital Angular Momentum from Wigner Distributions and 
Light-cone Wave Functions}

\author{C\'edric Lorc\'e}
\affiliation{IPN and LPT, Universit\'e Paris-Sud 11, 91405 Orsay, France}

\author{Barbara Pasquini}
\affiliation{Universit\`{a} degli Studi di Pavia, Dipartimento di Fisica Nucleare e Teorica}
\affiliation{INFN, Sezione di Pavia, Italy}

\author{Xiaonu Xiong}
\affiliation{Nuclear Science Division, Lawrence Berkeley National
Laboratory, Berkeley, CA 94720, USA}
\affiliation{Center for High-Energy
Physics, Peking University, Beijing 100871, China}

\author{Feng Yuan}
\affiliation{Nuclear Science Division, Lawrence Berkeley National Laboratory, Berkeley, CA 94720, USA}

\begin{abstract}
We investigate the quark orbital angular momentum of the nucleon in the
absence of gauge-field degrees of freedom, by using the concept of 
the Wigner distribution and the light-cone wave functions of the Fock
state expansion of the nucleon. The quark orbital angular momentum 
is obtained from the phase-space average of the orbital angular momentum 
operator weighted with the Wigner distribution of unpolarized quarks 
in a longitudinally polarized nucleon. We also derive the light-cone wave 
function representation of the orbital angular momentum. 
In particular, we perform an expansion in the nucleon Fock state space and 
decompose the orbital angular momentum into the $N$-parton state contributions. 
Explicit expressions are presented in terms of the light-cone wave functions
of the three-quark Fock state. Numerical results for the up and
down quark orbital angular momenta of the proton are shown
in the light-cone constituent quark model and the light-cone chiral 
quark-soliton model. 

\end{abstract}


\maketitle

\section{Introduction}
\label{sec:1}

The spin structure of the nucleon is one of the most important open questions in hadronic physics which has recently attracted an increasing attention. While the quark spin contribution is well known from the accurate measurements in polarized deep inelastic scattering~\cite{Airapetian:2007mh,Alexakhin:2006vx}, the situation is much more unclear for the remaining contributions from the quark orbital angular momentum (OAM) and those from the gluon spin and OAM. In the present paper we will focus on the quark OAM. As a first step, we ignore the contributions from the gauge-field degrees of freedom. Under this assumption, there is no ambiguity in the definition of the quark OAM operator~\cite{Jaffe:1989jz,Ji:1996ek,Burkardt:2008ua}. To unveil the underlying physics associated with the quark OAM, we discuss two different representations. The first one is obtained by using the concept of the Wigner distributions~\cite{Lorce:2011kd}, whereas the second one is based on light-cone wave functions (LCWFs). The extension to include the gauge-field degrees of freedom is left for a further study.

The Wigner distribution was originally constructed as the quantum mechanical analogue of the classical density operator in the phase space. Recently, it was also exploited to provide a multi-dimensional mapping of the partons in the nucleon~\cite{Ji:2003ak,Belitsky:2003nz,Lorce:2011kd}. An important aspect of the Wigner distribution is the possibility to calculate the expectation value of any physical observable from its phase-space average with the Wigner distribution as weighting factor. In particular, we will show that the quark OAM can be obtained from the Wigner distribution for unpolarized quarks in the longitudinally polarized nucleon.

On the other hand, the LCWFs provide the natural framework for the representation of OAM. This is due to the fact that the LCWFs are eigenstates of the total OAM for each $N$-parton configuration in the nucleon Fock space. In particular, we will consider the three valence-quark sector, giving the explicit representation of the quark OAM in terms of the partial-wave amplitudes of the nucleon state.

The plan of the paper is as follows. In Sec.~\ref{sec:2} we give the relevant definitions for the quark OAM operator and derive the expression of the OAM in terms of a Wigner distribution. In Sec.~\ref{sec:3} we first discuss the LCWF representation of the quark OAM, giving the decomposition into the contributions from the $N$-parton states. For the three-valence quark sector we further derive in Sec.~\ref{sec:4} the explicit expressions for the contributions from the different partial-wave amplitudes of the nucleon state to the OAM. The corresponding expressions for the distribution in $x$ of the OAM are collected in the Appendix. The formalism is finally applied in a light-cone constituent quark model and the light-cone version of the chiral quark-soliton model, and the corresponding results are discussed in Sec.~\ref{sec:5}. Concluding remarks are given in the last section.

\section{Quark orbital angular momentum}
\label{sec:2}
\subsection{Definitions}

If one neglects gauge-field degrees of freedom, the quark OAM 
operator for a given flavor $q$ can unambiguously be defined as
\be\label{OAMop1}
\widehat L^q_z\equiv-\frac{i}{2}\int\ud r^-\,\ud^2\uvec r\,\overline\psi^q(r^-,\uvec r)\gamma^+\left(\uvec r\times\!\stackrel{\leftrightarrow}{\uvec\nabla}_{\uvec r}\right)_z\psi^q(r^-,\uvec r),
\ee
where normal ordering is understood and the transverse vector $\uvec a=(a^1,a^2)$ has been introduced for the generic 4-vector $a$. Furthermore, in Eq.~\eqref{OAMop1} we used for convenience the notation $\stackrel{\leftrightarrow}{\uvec\nabla}\equiv\stackrel{\rightarrow}{\uvec\nabla}-\stackrel{\leftarrow}{\uvec\nabla}$. Using the Fourier transform of the quark field from the coordinate space to the momentum space
\be
\psi^q(r^-,\uvec r)=\int\frac{\ud k^+\,\ud^2\uvec k}{(2\pi)^3}\,e^{-i(k^+r^--\uvec k\cdot\uvec r)}\,\psi^q(k^+,\uvec k),
\ee
the quark OAM operator can alternatively be written as
\be\label{OAMop2}
\widehat L^q_z=-\frac{i}{2}\int\frac{\ud k^+\,\ud^2\uvec k}{(2\pi)^3}\,\overline\psi^q(k^+,\uvec k)\gamma^+\left(\uvec k\times\!\stackrel{\leftrightarrow}{\uvec\nabla}_{\uvec k}\right)_z\psi^q(k^+,\uvec k).
\ee
We define the quark OAM $\ell_z^q$ as the following matrix element of the quark OAM operator
\be
\label{oam}
\langle p',+|\widehat L^q_z|p,+\rangle\equiv\ell_z^q\,\langle p',+|p,+\rangle,
\ee
where the momenta of the incoming and outgoing nucleon are given by
\begin{equation}\label{symm-frame}
p=P-\tfrac{\Delta}{2}, \qquad p'=P+\tfrac{\Delta}{2}.
\end{equation}
Since the nucleon states $|p,\Lambda\rangle$ in Eq.~(\ref{oam}) are normalized as
\be
\langle p',+|p,+\rangle=2P^+\,\delta\left(\Delta^+\right)(2\pi)^3\,\delta^{(2)}\left(\uvec\Delta\right),
\ee
the forward limit $p'=p$ has to be treated with care. One can easily get rid of this normalization by integrating over $\Delta^+$ and $\uvec\Delta$
\be\label{OAMdef}
\ell_z^q=\int\frac{\ud\Delta^+}{2P^+}\,\frac{\ud^2\uvec\Delta}{(2\pi)^3}\,\langle p',+|\widehat L^q_z|p,+\rangle,
\ee
avoiding in this way to use normalizable wave packets or
infinite normalization factors.

\subsection{Orbital angular momentum from Wigner distributions}

In this section we derive the link between the quark OAM and the Wigner distributions. In particular, we show how it is possible to express Eq.~\eqref{OAMdef} as the phase-space average of the intuitive OAM operator $\left(\uvec r\times\uvec k\right)_z$ weighted with the Wigner distribution of unpolarized quarks in a longitudinally polarized nucleon.

From the definition of the OAM introduced in the previous section, it is natural to interpret the integrands of Eqs.~\eqref{OAMop1} and \eqref{OAMop2} as the quark OAM density operators in position and momentum space, respectively,
\begin{subequations}
\begin{align}
\widehat L^q_z(r^-,\uvec r)&=-\frac{i}{2}\,\overline\psi^q(r^-,\uvec r)\gamma^+\left(\uvec r\times\!\stackrel{\leftrightarrow}{\uvec\nabla}_{\uvec r}\right)_z\psi^q(r^-,\uvec r),\label{OAMopd1}\\
\widehat L^q_z(k^+,\uvec k)&=-\frac{i}{2}\,\overline\psi^q(k^+,\uvec k)\gamma^+\left(\uvec k\times\!\stackrel{\leftrightarrow}{\uvec\nabla}_{\uvec k}\right)_z\psi^q(k^+,\uvec k).\label{OAMopd2}
\end{align}
\end{subequations}
The generalization to the joint position and momentum space corresponds to the quark OAM density operator in the phase space
\be\label{OAM:ps}
\widehat L^q_z(r^-,\uvec r,k^+,\uvec k)=-\frac{i}{2}\int\frac{\ud z^-\,\ud^2\uvec z}{(2\pi)^3}\,e^{i(k^+z^--\uvec k\cdot\uvec z)}\,\overline\psi^q(r_f^-,\uvec r_f)\gamma^+\left(\uvec r\times\!\stackrel{\leftrightarrow}{\uvec\nabla}_{\uvec r}\right)_z\psi^q(r_i^-,\uvec r_i),
\ee
where $r_i=r+z/2$ and $r_f=r-z/2$. This operator contains the OAM density operators in position and momentum space as specific limits, \emph{i.e.}
\begin{subequations}
\begin{align}
\int\ud k^+\,\ud^2\uvec k\,\widehat L^q_z(r^-,\uvec r,k^+,\uvec k)&=\widehat L^q_z(r^-,\uvec r),\\
\int\ud r^-\,\ud^2\uvec r\,\widehat L^q_z(r^-,\uvec r,k^+,\uvec k)&=\widehat L^q_z(k^+,\uvec k).
\end{align}
\end{subequations}
Introducing the Wigner operator~\cite{Ji:2003ak,Belitsky:2003nz,Lorce:2011kd}
\be
\widehat W^{[\gamma^+]q}(r^-,\uvec r,k^+,\uvec k)=\frac{1}{2}\int\frac{\ud z^-\,\ud^2\uvec z}{(2\pi)^3}\,e^{i(k^+z^--\uvec k\cdot\uvec z)}\,\overline\psi^q(r_f^-,\uvec r_f)\gamma^+\psi^q(r_i^-,\uvec r_i),
\ee
the OAM density operator in the phase space can also be expressed as
\be
\widehat L^q_z(r^-,\uvec r,k^+,\uvec k)=2\left(\uvec r\times\uvec k\right)_z\,\widehat W^{[\gamma^+]q}(r^-,\uvec r,k^+,\uvec k).
\ee
Using the definition of the OAM in Eq.~\eqref{OAMdef}, we can write
\begin{align}
\ell_z^q&=\int\frac{\ud\Delta^+}{2P^+}\,\frac{\ud^2\uvec\Delta}{(2\pi)^3}\,\langle p',+|\int\ud r^-\,\ud^2\uvec r\,\ud k^+\,\ud^2\uvec k\,\widehat L^q_z(r^-,\uvec r,k^+,\uvec k)|p,+\rangle\nonumber\\
&=\int\frac{\ud\Delta^+}{P^+}\,\frac{\ud^2\uvec\Delta}{(2\pi)^3}\,\ud r^-\,\ud^2\uvec r\,\ud k^+\,\ud^2\uvec k\left(\uvec r\times\uvec k\right)_z\langle p',+|\widehat W^{[\gamma^+]q}(r^-,\uvec r,k^+,\uvec k)|p,+\rangle\nonumber\\
&=\int\frac{\ud\Delta^+}{P^+}\,\frac{\ud^2\uvec\Delta}{(2\pi)^3}\,\ud^2\uvec r\,\ud k^+\,\ud^2\uvec k\left(\uvec r\times\uvec k\right)_z\langle p',+|\widehat W^{[\gamma^+]q}(0,\uvec r,k^+,\uvec k)|p,+\rangle\int\ud r^-\,e^{i\Delta^+r^-}\nonumber\\
&=\int\ud x\,\ud^2\uvec k\,\ud^2\uvec r\left(\uvec r\times\uvec k\right)_z\,\rho^{[\gamma^+]q}_{++}(\uvec r,\uvec k,x),
\label{OAM:wigner}
\end{align}
where in the last line we introduced the Wigner distribution for unpolarized quark in a longitudinally polarized nucleon $\rho^{[\gamma^+]q}_{++}$:
\be
\rho^{[\Gamma]q}_{\Lambda'\Lambda}(\uvec r,\uvec k,x)=\int\frac{\ud^2\uvec\Delta}{(2\pi)^2}\,\langle P^+,\tfrac{\uvec\Delta}{2},\Lambda'|\widehat W^{[\Gamma]q}(0,\uvec r,xP^+,\uvec k)|P^+,-\tfrac{\uvec\Delta}{2},\Lambda\rangle
\ee
with $x=k^+/P^+$ the fraction of quark longitudinal momentum. Since we consider a nucleon state with its transverse center of momentum at the origin of the axes, we may identify the transverse coordinate $\uvec r$ with the impact parameter $\uvec b$~\cite{Burkardt:2000za,Burkardt:2002hr,Burkardt:2005td}.

In the same way as the impact-parameter dependent parton distributions are two-dimensional Fourier transforms of the generalized parton distributions (GPDs), the Wigner distributions are two-dimensional Fourier transforms of the so-called generalized transverse-momentum dependent parton distributions (GTMDs)~\cite{Lorce:2011kd,Lorce:2011dv,Meissner:2009ww}
\be
\rho^{[\Gamma]q}_{\Lambda'\Lambda}(\uvec b,\uvec k,x)=\int\frac{\ud^2\uvec\Delta}{(2\pi)^2}\,e^{-i\uvec\Delta\cdot\uvec b}\,W^{[\Gamma]q}_{\Lambda'\Lambda}(\uvec\Delta,\uvec k,x)
\label{wigner}
\ee
with the GTMD correlator given by
\be
\label{GTMD}
W^{[\Gamma]q}_{\Lambda'\Lambda}(\uvec\Delta,\uvec k,x)=\langle P^+,\tfrac{\uvec\Delta}{2},\Lambda'|\widehat W^{[\Gamma]q}(0,\uvec 0,xP^+,\uvec k)|P^+,-\tfrac{\uvec\Delta}{2},\Lambda\rangle
\ee
for the generic twist-2 Dirac matrix $\Gamma=\gamma^+,\gamma^+\gamma_5,i\sigma^{j+}\gamma_5$ with $j=1,2$. The hermiticity property of the GTMD correlator
\be
\left[W^{[\Gamma]q}_{\Lambda'\Lambda}(\uvec\Delta,\uvec k,x)\right]^*=
W^{[\Gamma]q}_{\Lambda\Lambda'}(-\uvec\Delta,\uvec k,x)
\ee
implies that $\rho^{[\gamma^+]q}_{++}$ is a real quantity, in accordance with its interpretation as a phase-space distribution.

Collecting the main results of this section, we can express the OAM in terms of Wigner distributions as
\be\label{OAMform}
\ell_z^q=\int\ud x\,\ud^2\uvec k\,\ud^2\uvec b\left(\uvec b\times\uvec k\right)_z\,\rho^{[\gamma^+]q}_{++}(\uvec b,\uvec k,x),
\ee
or, equivalently, in terms of GTMDs as~\cite{Lorce:2011kd} 
\begin{align}
\ell_z^q&=\int\ud x\,\ud^2\uvec k\left[i\left(\uvec k\times\uvec\nabla_{\uvec\Delta}\right)_z\,
W^{[\gamma^+]q}_{++}(\uvec\Delta,\uvec k,x)\right]_{\uvec\Delta=\uvec 0}\\
&=-\int\ud x\,\ud^2\uvec k\,\frac{\uvec k^2}{M^2}\,F^q_{1,4}(x,\uvec k^2,0,0),\label{lzform}
\end{align}
where $F^q_{1,4}$ follows the notation of Ref.~\cite{Meissner:2009ww}.

\section{LCWF overlap representation}
\label{sec:3}

Following the lines of Refs.~\cite{Diehl:2000xz,Brodsky:2000xy}, we can obtain an overlap representation of Eq.~\eqref{lzform} in terms of LCWFs. Since the quark OAM operator is diagonal in light-cone helicity, flavor and color spaces, we can write the quark OAM as the sum of the contributions from the $N$-parton Fock states
\be
\ell_z^q=\sum_{N,\beta}\ell_z^{N\beta,q},
\label{overlap:l}
\ee
where
\be\label{lzDelta}
\ell_z^{N\beta,q}=i\int\left[\ud x\right]_N\left[\ud^2\uvec k\right]_N\sum_{i=1}^N\delta_{qq_i}\left[\left(\uvec k_i\times\uvec\nabla_{\uvec\Delta}\right)_z\,\Psi^{*+}_{N\beta}(r^{out})\,\Psi^+_{N\beta}(r^{in})\right]_{\uvec\Delta=\uvec 0}.
\ee
Note that this expression is consistent with the wave-packet approach of Ref.~\cite{Hagler:2003jw}, except for the difference by a factor of $1/2$ which follows from different conventions for the light-cone components. In Eq.~\eqref{lzDelta}, the integration measures are given by
\begin{align}
\left[\ud x\right]_N&=\left[\prod_{i=1}^N\ud x_i\right]\delta\left(1-\sum_{i=1}^N x_i\right),\\
\left[\ud^2\uvec k\right]_N&=\left[\prod_{i=1}^N\frac{\ud^2\uvec k_i}{2(2\pi)^3}\right]2(2\pi)^3\,\delta^{(2)}\left(\sum_{i=1}^N\uvec k_i\right),
\end{align}
and the $N$-parton LCWF $\Psi^\Lambda_{N\beta}(r)$ depends on the hadron light-cone helicity $\Lambda$, on the Fock-state composition $\beta$ denoting collectively quark light-cone helicities, flavors and colors, and on the relative quark momenta collectively denoted by $r$. In particular, for the active parton $i$ one has
\begin{align}
x^{in}_i&=x_i,&\uvec k^{in}_i&=\uvec k_i-\left(1-x_i\right)\tfrac{\uvec\Delta}{2},\\
x^{out}_i&=x_i,&\uvec k^{out}_i&=\uvec k_i+\left(1-x_i\right)\tfrac{\uvec\Delta}{2},
\intertext{and for the the spectator partons $j\neq i$, one has}
x^{in}_j&=x_j,&\uvec k^{in}_j&=\uvec k_j+x_j\,\tfrac{\uvec\Delta}{2},\\
x^{out}_j&=x_j,&\uvec k^{out}_j&=\uvec k_j-x_j\,\tfrac{\uvec\Delta}{2}.
\end{align}

Acting with $\uvec\nabla_{\uvec\Delta}$ on the LCWFs in Eq.~\eqref{lzDelta} and then setting $\uvec\Delta=\uvec 0$ leads to the following expression\footnote{We find the same expression starting from the representation of the quark OAM operator in momentum space given by Eq.~\eqref{OAMop2} and integrating over $\uvec\Delta$ to kill the $\delta$ function subjected to $\uvec\nabla_{\uvec k}$.}
\be\label{OR}
\ell_z^{N\beta,q}=-\frac{i}{2}\int\left[\ud x\right]_N\left[\ud^2\uvec k\right]_N\sum_{i=1}^N\delta_{qq_i}\sum_{n=1}^N\left(\delta_{ni}-x_n\right)
\left[\Psi^{*+}_{N\beta}(r)\left(\uvec k_i\times\stackrel{\leftrightarrow}{\uvec\nabla}_{\uvec k_n}\right)_z\Psi^+_{N\beta}(r)\right].
\ee
This expression coincides with our intuitive picture of OAM. Indeed, since the transverse position $\uvec r$ is represented in transverse-momentum space by $i\uvec\nabla_{\uvec k}$, it means that
\be
i\sum_{n=1}^N\left(\delta_{ni}-x_n\right)\uvec\nabla_{\uvec k_n}\sim\uvec r_i-\sum_{n=1}^Nx_n\,\uvec r_n=\uvec b_i
\ee
represents the transverse position of the active quark \emph{relative} to the transverse center of momentum $\uvec R=\sum_nx_n\,\uvec r_n$. Moreover, by construction, $\uvec k_i$ represents the relative transverse momentum of the active quark. It follows that Eq.~\eqref{OR} gives the \emph{intrinsic} quark OAM defined with respect to the transverse center of momentum\footnote{This is completely analogous to non-relativistic mechanics where the intrinsic OAM is defined relative to the center of mass
\begin{align*}
\sum_n\vec r_n\times\vec p_n&=\sum_n(\vec r_n-\vec R)\times(\vec p_n-x_n\vec P)+\vec R\times\sum_n(\vec p_n-x_n\vec P)+\sum_nx_n\vec r_n\times\vec P\\
&=\sum_i\vec r^{\phantom{,}rel}_i\times\vec p^{\phantom{,}rel}_i+\vec R\times\vec P,
\end{align*}
with $\vec R=\sum_nx_n\vec r_n$, $\vec P=\sum_n\vec p_n$, $x_n=m_n/M$ and $M=\sum_nm_n$. This analogy comes from the galilean symmetry of transverse space on the light cone.}.

An important property of the LCWFs is that they are eigenstates of the \emph{total} OAM
\be
-i\sum_{n=1}^N\left(\uvec k_n\times\uvec\nabla_{\uvec k_n}\right)_z\Psi^\Lambda_{N\beta}(r)=l_z\,\Psi^\Lambda_{N\beta}(r)
\ee
with eigenvalue $l_z=(\Lambda-\sum_n\lambda_n)/2$. As a consequence, from the overlap representation \eqref{OR} it is straightforward to show that the total OAM is given by
\be\label{rel}
\ell_z=\sum_{N,l_z}l_z\,\rho_{Nl_z},
\ee
where 
\be\label{norm}
\rho_{Nl_z}\equiv\sum_{\beta'}\delta_{l_zl'_z}\int\left[\ud x\right]_N\left[\ud^2\uvec k\right]_N\left|\Psi^+_{N\beta'}(r)\right|^2
\ee
is the probability to find the nucleon with light-cone helicity $\Lambda=+$ in an $N$-parton state with eigenvalue $l_z$ of the total OAM.

\section{Partial-wave decomposition of quark orbital angular momentum}
\label{sec:4}

In this section we give the explicit expressions for the OAM $\ell^q_z$ in terms of the different partial-wave components of the three-quark state of the nucleon.

Working in the so-called ``uds'' basis~\cite{Franklin:1968pn,Capstick:1986bm}, the proton state is given in terms of a completely symmetrized wave function of the form
\be\label{eq:2}
|P,+\rangle=|P,+\rangle_{uud}+|P,+\rangle_{udu}+|P,+\rangle_{duu} \,.
\ee
In this symmetrization, the state $|P,+\rangle_{udu}$ is obtained from $|P,+\rangle_{uud}$ by interchanging the second and third spin and space coordinates as well as the indicated quark type, with a similar interchange of the first and third coordinates for $|P,+\rangle_{duu}$. For the calculation of matrix elements of one-body operators, as in the case of the OAM operator, it is sufficient to use only the $uud$ order. As outlined in the previous section, the LCWF $\Psi^{\Lambda;uud}_{\lambda_1\lambda_2\lambda_3}$ of the $uud$ component is eigenstate of the total OAM, with eigenvalues $l_z=(\Lambda-\sum_{i=1}^3\lambda_i)/2$. In particular, for a nucleon with helicity $\Lambda=+$ we can have four partial waves with $l_z=0,\pm 1,2$ corresponding to combinations of $\Psi^{+;uud}_{\lambda_1\lambda_2\lambda_3}$ with appropriate values of the quark helicities $\lambda_i$. However, the $16$ helicity configurations of $\Psi^{\Lambda;uud}_{\lambda_1\lambda_2\lambda_3}$ are not all independent. Parity and isospin symmetries leave only $6$ independent functions $\psi^{(i)}$ of quark momenta. In particular, the complete three-quark light-cone Fock expansion has the following structure~\cite{Ji:2002xn,Burkardt:2002uc,Ji:2003yj}:
\be\label{eq:Fock}
|P,+\rangle=|P,+\rangle^{l_z=0}+|P,+\rangle^{l_z=1}+|P,+\rangle^{l_z=-1}+|P,+\rangle^{l_z=2},
\ee
where
\begin{subequations}
\begin{align}
|P,+\rangle^{l_z=0}&=\int\frac{\left[\ud x\right]_3\left[\ud^2\uvec k\right]_3}{\sqrt{x_1x_2x_3}}\left[\psi^{(1)}(1,2,3)+i\epsilon^{\alpha\beta}k_{1\alpha}k_{2\beta}\,\psi^{(2)}(1,2,3)\right] \nonumber\\
&\quad\times\frac{\epsilon^{ijk}}{\sqrt{6}}\,u^{\dagger}_{i+}(1)\left(u^{\dagger}_{j-}(2)d^{\dagger}_{k+}(3)-d^{\dagger}_{j-}(2)u^{\dagger}_{k+}(3)\right)|0\rangle, \label{lca1}\\
|P,+\rangle^{l_z=1}&=\int\frac{\left[\ud x\right]_3\left[\ud^2\uvec k\right]_3}{\sqrt{x_1x_2x_3}}\left[k_{1R}\,\psi^{(3)}(1,2,3)+k_{2R}\,\psi^{(4)}(1,2,3)\right]\nonumber\\
&\quad\times\frac{\epsilon^{ijk}}{\sqrt{6}}\left(u^{\dagger}_{i+}(1)u^{\dagger}_{j-}(2)d^{\dagger}_{k-}(3)-d^{\dagger}_{i+}(1)u^{\dagger}_{j-}(2)u^{\dagger}_{k-}(3)\right)|0\rangle,\label{lca2}\\
|P,+\rangle^{l_z=-1}&=\int\frac{\left[\ud x\right]_3\left[\ud^2\uvec k\right]_3}{\sqrt{x_1x_2x_3}}\,(-k_{2L})\,\psi^{(5)}(1,2,3)\nonumber \\
&\quad\times\frac{\epsilon^{ijk}}{\sqrt{6}}\,u^{\dagger}_{i+}(1)\left(u^{\dagger}_{j+}(2)d^{\dagger}_{k+}(3)-d^{\dagger}_{j+}(2)u^{\dagger}_{k+}(3)\right)|0\rangle,\label{lca3}\\
|P,+\rangle^{l_z=2}&=\int\frac{\left[\ud x\right]_3\left[\ud^2\uvec k\right]_3}{\sqrt{x_1x_2x_3}}\,k_{1R}k_{3R}\,\psi^{(6)}(1,2,3)\nonumber\\
&\quad\times\frac{\epsilon^{ijk}}{\sqrt{6}}\,u^{\dagger}_{i-}(1)\left(d^{\dagger}_{j-}(2)u^{\dagger}_{k-}(3)-u^{\dagger}_{j-}(2)d^{\dagger}_{k-}(3)\right)|0\rangle.\label{lca4}
\end{align}
\end{subequations}
In Eqs.~\eqref{lca1}-\eqref{lca4} $\alpha,\beta=1,2$ are transverse indices, $u^\dagger_{i\lambda}$ ($u_{i\lambda}$) and $d^\dagger_{i\lambda}$ ($d_{i\lambda}$) are creation (annihilation) operators of $u$ and $d$ quarks with helicity $\lambda$ and color $i$, and $\psi^{(j)}$ are functions of quark momenta, with argument $i$ representing $x_i$ and $\uvec k_i$ and with a dependence on the transverse momenta of the form $\uvec k_i\cdot\uvec k_j$ only. We also used the notation $k_{R,L}=k_x\pm ik_y$.

The relations between the wave function amplitudes $\psi^{(i)}$ and the three-quark LCWFs $\Psi^{+;uud}_{\lambda_1\lambda_2\lambda_3}$ are given by
\begin{subequations}
\begin{align}
\Psi^{+;uud}_{+-+}&=\psi^{(1,2)}(1,2,3),\\
\Psi^{+;uud}_{-++}&=\psi^{(1,2)}(2,1,3),\\
\Psi^{+;uud}_{++-}&=-\left[\psi^{(1,2)}(1,3,2)+\psi^{(1,2)}(2,3,1)\right],\\
\Psi^{+;uud}_{+--}&=\psi^{(3,4)}(1,2,3),\\
\Psi^{+;uud}_{-+-}&=\psi^{(3,4)}(2,1,3),\\
\Psi^{+;uud}_{--+}&=-\left[\psi^{(3,4)}(1,3,2)+\psi^{(3,4)}(2,3,1)\right],\\
\Psi^{+;uud}_{+++}&=-k_{2L}\,\psi^{(5)}(1,2,3),\\
\Psi^{+;uud}_{---}&=-k_{1R}k_{3R}\,\psi^{(6)}(1,2,3),
\end{align}
\end{subequations}
where we defined
\begin{subequations}
\begin{align}
\psi^{(1,2)}(1,2,3)&=\left[\psi^{(1)}(1,2,3)+i\epsilon^{\alpha\beta}k_{1\alpha}k_{2\beta}\,\psi^{(2)}(1,2,3)\right],\\
\psi^{(3,4)}(1,2,3)&=\left[k_{1R}\psi^{(3)}(1,2,3)+k_{2R}\psi^{(4)}(1,2,3)\right].
\end{align}
\end{subequations}
The corresponding expressions for hadron helicity $\Lambda=-$ are obtained thanks to light-cone parity symmetry
\be
\Psi^{-\Lambda;uud}_{-\lambda_1-\lambda_2-\lambda_3}(\{x_i,k_{ix},k_{iy}\})=-\Psi^{\Lambda;uud}_{\lambda_1\lambda_2\lambda_3}(\{x_i,-k_{ix},k_{iy}\}).
\ee

Using the partial-wave decomposition of the nucleon state in Eqs.~\eqref{lca1}-\eqref{lca4} we can seperately calculate the results for $\ell_z^{q,l_z}$ corresponding to the contribution of the the quark with flavor $q$ in the Fock state component with OAM $l_z$. Summing over all flavors we find, in agreement with Eq.~\eqref{rel}, $\ell_z^{l_z}=l_z\,\rho_{l_z}$ (we omit the index $N=3$ since we considered only the three-quark Fock contribution) with $\rho_{l_z}=\,^{l_z}\!\langle P,+|P,+\rangle^{l_z}$. In the Appendix we also give the results for the partial-wave decomposition of the distribution in $x$ of the OAM for the $u$- and $d$-quark contributions.

For the $l_z=0$ component, we find:
\\
${\bullet}$ for total $u$-quark contribution
\begin{align}
&\ell_z^{u,l_z=0}=\int\left[\ud x\right]_3\left[\ud^2\uvec k\right]_3\Big\{\Big[(1-x_1)\uvec k_1\cdot \uvec k_2+x_2\uvec k_1^{\, 2}\Big]\nonumber\\
&\times\Big[-\psi^{(1)}(1,2,3)\psi^{(2)}(1,2,3)+\psi^{(1)}(2,1,3)\psi^{(2)}(2,1,3)\Big]\nonumber\\
&-\Big[\psi^{(1)}(1,2,3)+\psi^{(1)}(3,2,1)\Big]\nonumber\\
&\times\Big[\Big((1-x_1)\uvec k_1\cdot\uvec k_2+x_2\uvec k_1^{\, 2}\Big)\psi^{(2)}(1,2,3)+\Big(x_2\uvec k_1\cdot\uvec k_3-x_3\uvec k_1\cdot\uvec k_2\Big)\psi^{(2)}(3,2,1)\Big]\nonumber\\
&+(\uvec k_1\times\uvec k_2)\Big[\psi^{(1)}(1,2,3)\Big(\uvec k_1\times\widetilde{\uvec\nabla}_1+\uvec k_2\times\widetilde{\uvec\nabla}_2\Big)\psi^{(2)}(1,2,3)\nonumber\\
&-\psi^{(2)}(1,2,3)\Big(\uvec k_1\times\widetilde{\uvec\nabla}_1+\uvec k_2\times\widetilde{\uvec\nabla}_2\Big)\psi^{(1)}(1,2,3)\nonumber\\
&-\psi^{(1)}(1,3,2)\Big(\uvec k_1\times\widetilde{\uvec\nabla}_1+\uvec k_2\times\widetilde{\uvec\nabla}_2\Big)\Big(\psi^{(2)}(1,3,2)-\psi^{(2)}(2,3,1)\Big)\nonumber\\
&+\psi^{(2)}(1,3,2)\Big(\uvec k_1\times\widetilde{\uvec\nabla}_1+\uvec k_2\times\widetilde{\uvec\nabla}_2\Big)\Big(\psi^{(1)}(1,3,2)+\psi^{(1)}(2,3,1)\Big)\Big]\Big\}\ ;
\label{lzt0u}
\end{align}
\\
${\bullet}$ for total $d$-quark contribution
\begin{align}
&\ell_z^{d,l_z=0}=\int\left[\ud x\right]_3\left[\ud^2\uvec k\right]_3\Big\{\Big[x_1\uvec k_2\cdot\uvec k_3-x_2\uvec k_1\cdot\uvec k_3\Big]\,\psi^{(1)}(1,2,3)\psi^{(2)}(1,2,3)\nonumber\\
&+\Big[(1-x_2)\uvec k_1\cdot\uvec k_2+x_1\uvec k_2^{\, 2}\Big]\,\Big[\psi^{(1)}(1,2,3)+\psi^{(1)}(3,2,1)\Big]\psi^{(2)}(1,2,3)\,\nonumber\\
&+(\uvec k_1\times\uvec k_2)\Big[\psi^{(1)}(1,2,3)\Big(\uvec k_3\times\widetilde{\uvec\nabla}_3\Big)\psi^{(2)}(1,2,3)\nonumber\\
&-\psi^{(2)}(1,2,3)\Big(\uvec k_3\times\widetilde{\uvec\nabla}_3\Big)\psi^{(1)}(1,2,3)\nonumber\\
&-\psi^{(1)}(1,3,2)\Big(\uvec k_3\times\widetilde{\uvec\nabla}_3\Big)\Big(\psi^{(2)}(1,3,2)-\psi^{(2)}(2,3,1)\Big)\nonumber\\
&+\psi^{(2)}(1,3,2)\Big(\uvec k_3\times\widetilde{\uvec\nabla}_3\Big)\Big(\psi^{(1)}(1,3,2)+\psi^{(1)}(2,3,1)\Big)\Big]\Big\}\ .
\label{lzt0d}
\end{align}
In Eq.~(\ref{lzt0u}) we used the following definition
\be
\widetilde{\uvec\nabla}_1\psi^{(i)}(1,2,3)=\left[(1-x_1)\uvec\nabla_{\uvec k_1}-x_2\uvec\nabla_{\uvec k_2}\right]\psi^{(i)}(1,2,\uvec k_3=-\uvec k_1-\uvec k_2)\ ,
\ee
and similarly for $\widetilde{\uvec\nabla}_2$. Furthermore, in Eq.~(\ref{lzt0d}) the operator $\widetilde{\uvec\nabla}_3$ is defined as
\be
\widetilde{\uvec\nabla}_3\psi^{(i)}(1,2,3)=-\left(x_1\uvec\nabla_{\uvec k_1}+x_2\uvec\nabla_{\uvec k_2}\right)\psi^{(i)}(1,2,\uvec k_3=-\uvec k_1-\uvec k_2)\ .
\ee
Using the momentum conservation constraint $\uvec k_1+\uvec k_2+\uvec k_3=\uvec 0$, one finds $\ell_z^{u,l_z=0}=-\ell_z^{d,l_z=0}$. One then recovers the fact that the total contribution from the $l_z=0$ component is equal to zero
\be
\ell_z^{l_z=0}=0.
\ee

For the $l_z=1$ component, we find:
\\
${\bullet}$ for the total $u$-quark contribution
\begin{align}
&\ell_z^{u,l_z=1}=\int\left[\ud x\right]_3\left[\ud^2\uvec k\right]_3\Big\{\Big[\uvec k_1\cdot\uvec k_2\psi^{(3)}(1,2,3)+\uvec k_2^{\, 2}\psi^{(4)}(1,2,3)\Big]\nonumber\\
&\times\Big[-2x_1\psi^{(3)}(1,2,3)+2(1-x_2)\psi^{(4)}(1,2,3)-x_1\psi^{(3)}(1,3,2)-x_3\psi^{(4)}(1,3,2)\Big]\nonumber\\
&+\Big[\uvec k_1^{\, 2}\psi^{(3)}(1,2,3)) +\uvec k_1\cdot \uvec k_2\psi^{(4)}(1,2,3)\Big]\Big[(1-x_1)\psi^{(3)}(1,2,3)-x_2\psi^{(4)}(1,2,3)\Big]\nonumber\\
&-\Big[\uvec k_1\cdot\uvec k_3\psi^{(3)}(1,2,3)+\uvec k_2\cdot\uvec k_3\psi^{(4)}(1,2,3)\Big]\nonumber\\
&\times\Big[x_1\psi^{(3)}(1,2,3)+x_2\psi^{(4)}(1,2,3)+x_1\psi^{(3)}(1,3,2)-(1-x_3)\psi^{(4)}(1,3,2)\Big]\nonumber\\
&-(\uvec k_1\times\uvec k_2)\psi^{(3)}(1,2,3)\,\Big[\Big(\uvec k_2\times\widetilde{\uvec\nabla}_2+\uvec k_3\times\widetilde{\uvec\nabla}_3\Big)\psi^{(4)}(1,3,2)-\Big(\uvec k_2\times\widetilde{\uvec\nabla}_2\Big)\psi^{(4)}(1,2,3)\Big]\nonumber\\
&-(\uvec k_1\times\uvec k_2)\psi^{(4)}(1,2,3)\,\Big[\Big(\uvec k_2\times\widetilde{\uvec\nabla}_2+\uvec k_3\times\widetilde{\uvec\nabla}_3\Big)\Big(\psi^{(3)}(1,3,2)-\psi^{(4)}(1,3,2)\Big)\nonumber\\
&-\Big(\uvec k_2\times\widetilde{\uvec\nabla}_2\Big)\psi^{(3)}(1,2,3)\Big]\Big\}\ ;
\label{eq:lz1u}
\end{align}
\\
${\bullet}$ for the total $d$-quark contribution
\begin{align}
&\ell_z^{d,l_z=1}=\int\left[\ud x\right]_3\left[\ud^2\uvec k\right]_3\Big\{\Big[\uvec k_1^{\, 2}\psi^{(3)}(1,2,3))+\uvec k_1\cdot \uvec k_2\psi^{(4)}(1,2,3)\Big]\nonumber\\
&\times\Big[(1-x_1)\psi^{(3)}(1,2,3)-x_2\psi^{(4)}(1,2,3)+(1-x_{1})\psi^{(3)}(1,3,2)-x_3\psi^{(4)}(1,3,2)\Big]\nonumber\\
&-\Big[\uvec k_1\cdot\uvec k_3\psi^{(3)}(1,2,3)+\uvec k_2\cdot\uvec k_3\psi^{(4)}(1,2,3)\Big]\Big[x_1\psi^{(3)}(1,2,3)+x_2\psi^{(4)}(1,2,3)\Big]\nonumber\\
&-(\uvec k_1\times\uvec k_2)\psi^{(3)}(1,2,3)\,\Big[\Big(\uvec k_1\times\widetilde{\uvec\nabla}_1\Big)\psi^{(4)}(1,3,2)-\Big(\uvec k_1\times\widetilde{\uvec\nabla}_1+\uvec k_3\times\widetilde{\uvec\nabla}_3\Big)\psi^{(4)}(1,2,3)\Big]\nonumber\\
&-(\uvec k_1\times\uvec k_2)\psi^{(4)}(1,2,3)\,\Big[\Big(\uvec k_1\times\widetilde{\uvec\nabla}_1\Big)\Big(\psi^{(3)}(1,3,2)-\psi^{(4)}(1,3,2)\Big)\nonumber\\
&+\Big(\uvec k_1\times\widetilde{\uvec\nabla}_1+\uvec k_3\times\widetilde{\uvec\nabla}_3\Big)\psi^{(3)}(1,2,3)\Big]\Big\}\ .
\label{eq:lz1d}
\end{align}
The sum of the $u$ and $d$ contributions from the $l_z=1$ component gives
\begin{align}
\ell_z^{l_z=1}=&\int\left[\ud x\right]_3\left[\ud^2\uvec k\right]_3\nonumber\\
&\times\Big\{\Big[2\psi^{(3)}(1,2,3)+\psi^{(3)}(1,3,2)\Big]\Big[\uvec k_1^{\, 2}\psi^{(3)}(1,2,3)+\uvec k_1\cdot\uvec k_2\psi^{(4)}(1,2,3)\Big]\nonumber\\
&+2\psi^{(4)}(1,2,3)\Big[\uvec k_1\cdot\uvec k_2\psi^{(3)}(1,2,3)-\uvec k_2^{\, 2}\psi^{(4)}(1,2,3)\Big]\nonumber\\
&+\psi^{(4)}(1,3,2)\Big[\uvec k_1\cdot\uvec k_3\psi^{(3)}(1,2,3)+\uvec k_2\cdot\uvec k_3\psi^{(4)}(1,2,3)\Big]\Big\}\nonumber\\
=&\,\,\rho_{l_z=1}.
\end{align}

For $l_z=-1$ component, we find:
\\
${\bullet}$ for the total $u$-quark contribution
\begin{align}
&\ell_z^{u,l_z=-1}=\int\left[\ud x\right]_3\left[\ud^2\uvec k\right]_3\Big\{\Big[\uvec k_1\cdot\uvec k_2\psi^{(5)}(1,2,3)-\uvec k_1\cdot\uvec k_3\psi^{(5)}(1,3,2)\Big]\nonumber\\
&\times\Big[x_2\psi^{(5)}(1,2,3)-x_3\psi^{(5)}(1,3,2)-x_3\psi^{(5)}(2,3,1)-(1-x_1)\psi^{(5)}(2,1,3)\Big]\nonumber\\
&+\Big[\uvec k_2^2\psi^{(5)}(1,2,3)-\uvec k_2\cdot\uvec k_3\psi^{(5)}(1,3,2)\Big]\nonumber\\
&\times\Big[x_1\psi^{(5)}(2,1,3)-x_3\psi^{(5)}(1,3,2)-x_3\psi^{(5)}(2,3,1)-(1-x_2)\psi^{(5)}(1,2,3)\Big]\nonumber\\
&+(\uvec k_2\times\uvec k_3)\Big[\psi^{(5)}(1,3,2)\Big(\uvec k_1\times\widetilde{\uvec\nabla}_1+\uvec k_2\times\widetilde{\uvec\nabla}_2\Big)\Big(\psi^{(5)}(2,1,3)-\psi^{(5)}(1,2,3)\Big)\nonumber\\
&+\psi^{(5)}(1,2,3)\Big(\uvec k_1\times\widetilde{\uvec\nabla}_1+\uvec k_2\times\widetilde{\uvec\nabla}_2\Big)\Big(\psi^{(5)}(1,3,2)+\psi^{(5)}(2,1,3)+\psi^{(5)}(2,3,1)\Big)\Big]\Big\}\ ;
\end{align}
\\
${\bullet}$ for the total $d$-quark contribution
\begin{align}
&\ell_z^{d,l_z=-1}=\int\left[\ud x\right]_3\left[\ud^2\uvec k\right]_3\Big\{\Big[\uvec k_2\cdot\uvec k_3\psi^{(5)}(1,2,3)-\uvec k_3^{\, 2}\psi^{(5)}(1,3,2)\Big]\nonumber\\
&\times\Big[x_1\psi^{(5)}(2,1,3)+x_2\psi^{(5)}(1,2,3)+(1-x_3)\Big(\psi^{(5)}(1,3,2)+\psi^{(5)}(2,3,1)\Big)\Big]\nonumber\\
&+(\uvec k_2\times\uvec k_3)\Big[\psi^{(5)}(1,3,2)\Big(\uvec k_3\times\widetilde{\uvec\nabla}_3\Big)\Big(\psi^{(5)}(2,1,3)-\psi^{(5)}(1,2,3)\Big)\nonumber\\
&+\psi^{(5)}(1,2,3)\Big(\uvec k_3\times\widetilde{\uvec\nabla}_3\Big)\Big(\psi^{(5)}(1,3,2)+\psi^{(5)}(2,1,3)+\psi^{(5)}(2,3,1)\Big)\Big]\Big\}\ .
\end{align}
Adding the $u$ and $d$ contributions, we find that the
$l_z=-1$ partial wave contributes to the total orbital angular momentum as
\begin{align}
\ell_z^{l_z=-1}=&\int\left[\ud x\right]_3\left[\ud^2\uvec k\right]_3\Big\{-\Big[\uvec k_1\cdot \uvec k_2\psi^{(5)}(1,2,3)-\uvec k_1\cdot\uvec k_3\psi^{(5)}(1,3,2)\Big]\psi^{(5)}(2,1,3)\nonumber\\
&-\Big[\uvec k_2^2\psi^{(5)}(1,2,3)-\uvec k_2\cdot\uvec k_3\psi^{(5)}(1,3,2)\Big]\psi^{(5)}(1,2,3)\nonumber\\
&+\Big[\uvec k_2\cdot\uvec k_3\psi^{(5)}(1,2,3)-\uvec k_3^{\, 2}\psi^{(5)}(1,3,2)\Big]\Big[\psi^{(5)}(1,3,2)+\psi^{(5)}(2,3,1)\Big]\Big\}\nonumber\\
=&-\rho_{l_z=-1}.
\end{align}

For $l_z=2$ component, we find:
\\${\bullet}$ for the total $u$-quark contribution
\begin{align}
&\ell_z^{u,l_z=2}=\int\left[\ud x\right]_3\left[\ud^2\uvec k\right]_3\Big\{\uvec k_1^{\, 2}\Big[(1-x_1)\uvec k_3^{\, 2}-x_3\uvec k_1\cdot\uvec k_3\Big]\nonumber\\
&\times\Big[\psi^{(6)}(1,2,3)\Big(2\psi^{(6)}(1,2,3)+\psi^{(6)}(3,2,1)\Big)+\psi^{(6)}(3,2,1)\Big(\psi^{(6)}(1,2,3)+\psi^{(6)}(3,2,1)\Big)\Big]\nonumber\\
&-\uvec k_1^{\, 2}\Big[x_2\uvec k_3^{\, 2}+x_3\uvec k_2\cdot\uvec k_3\Big]\nonumber\\
&\times\Big[\psi^{(6)}(1,2,3)\Big(\psi^{(6)}(2,1,3)-\psi^{(6)}(3,1,2)\Big)-\psi^{(6)}(3,2,1)\psi^{(6)}(3,1,2)\Big]\nonumber\\
&-\uvec k_1^{\, 2}\Big[(1-x_1)\uvec k_2\cdot\uvec k_3-x_2\uvec k_1\cdot\uvec k_3\Big]\nonumber\\
&\times\Big[\psi^{(6)}(1,2,3)\Big(2\psi^{(6)}(1,3,2)+\psi^{(6)}(2,3,1)\Big)+\psi^{(6)}(3,2,1)\psi^{(6)}(1,3,2)\Big]\nonumber\\
&-\Big[x_1\uvec k_1\cdot\uvec k_2\uvec k_3^{\,2}+x_3\uvec k_2\cdot\uvec k_3\uvec k_1^{\, 2}\Big]\psi^{(6)}(1,2,3)\psi^{(6)}(1,2,3)\nonumber\\
&-\Big[(1-x_2)\uvec k_2\cdot\uvec k_3\uvec k_1^{\, 2}-x_1(2\uvec k_1\cdot\uvec k_2\uvec k_2\cdot\uvec k_3-\uvec k_1\cdot\uvec k_3\uvec k_2^{\, 2})\Big]\nonumber\\
&\times\Big[\psi^{(6)}(1,2,3)\Big(\psi^{(6)}(1,3,2)+\psi^{(6)}(2,3,1)\Big)-\psi^{(6)}(3,2,1)\psi^{(6)}(2,3,1)\Big]\nonumber\\
&+(\uvec k_1\times\uvec k_2)\psi^{(6)}(1,2,3)\Big[\uvec k_3^{\, 2}\Big(\uvec k_1\times\widetilde{\uvec\nabla}_1+\uvec k_2\times\widetilde{\uvec\nabla}_2\Big)\psi^{(6)}(2,1,3)\nonumber\\
&-\uvec k_3^{\, 2}\Big(\uvec k_1\times\widetilde{\uvec\nabla}_1+\uvec k_3\times\widetilde{\uvec\nabla}_3\Big)\psi^{(6)}(3,1,2)\nonumber\\
&+\uvec k_1^{\, 2}\Big(\uvec k_1\times\widetilde{\uvec\nabla}_1\Big)\psi^{(6)}(1,3,2)+\uvec k_1^{\, 2}\Big(\uvec k_1\times\widetilde{\uvec\nabla}_1+\uvec k_2\times\widetilde{\uvec\nabla}_2\Big)\psi^{(6)}(2,3,1)\Big]\Big\}\ ;
\end{align}
\\
${\bullet}$ for the total $d$-quark contribution
\begin{align}
&\ell_z^{d,l_z=2}=\int\left[\ud x\right]_3\left[\ud^2\uvec k\right]_3\Big\{-\Big[x_1\uvec k_1\cdot\uvec k_2\uvec k_3^{\, 2}+x_3\uvec k_2\cdot\uvec k_3\uvec k_1^{\, 2}\Big]\nonumber\\
&\psi^{(6)}(1,2,3)\Big[\psi^{(6)}(1,2,3)+\psi^{(6)}(3,2,1)\Big]\nonumber\\
&-\Big[(1-x_2)\uvec k_1\cdot\uvec k_2\uvec k_3^{\, 2}-x_3(2\uvec k_1\cdot\uvec k_2\uvec k_2\cdot\uvec k_3-\uvec k_1\cdot\uvec k_3\uvec k_2^{\,2})\Big]\nonumber\\
&\times\psi^{(6)}(3,1,2)\Big[\psi^{(6)}(1,2,3)+\psi^{(6)}(3,2,1)\Big]\nonumber\\
&-\uvec k_3^{\, 2}\Big[x_2\uvec k_1\cdot\uvec k_3-(1-x_3)\uvec k_1\cdot\uvec k_2\Big]\psi^{(6)}(1,2,3)\psi^{(6)}(2,1,3)\nonumber\\
&-\uvec k_3^{\,2}\Big[x_1\uvec k_1\cdot\uvec k_3-(1-x_3)\uvec k_1^{\, 2}\Big]\psi^{(6)}(1,2,3)\psi^{(6)}(1,2,3)\nonumber\\
&+\uvec k_3^{\,2}\Big[x_1\uvec k_1\cdot\uvec k_2+x_2\uvec k_1^{\, 2}\Big]\psi^{(6)}(1,2,3)\Big[\psi^{(6)}(1,3,2)+\psi^{(6)}(2,3,1)\Big]\nonumber\\
&-(\uvec k_1\times\uvec k_2)\psi^{(6)}(1,2,3)\Big[\uvec k_3^{\, 2}\Big(\uvec k_2\times\widetilde{\uvec\nabla}_2\Big)\psi^{(6)}(3,1,2)+\uvec k_3^{\, 2}\Big(\uvec k_3\times\widetilde{\uvec\nabla}_3\Big)\psi^{(6)}(2,1,3)\nonumber\\
&-\uvec k_1^{\, 2}\Big(\uvec k_2\times\widetilde{\uvec\nabla}_2+\uvec k_3\times\widetilde{\uvec\nabla}_3\Big)\psi^{(6)}(1,3,2)+\uvec k_1^{\, 2}\Big(\uvec k_3\times\widetilde{\uvec\nabla}_3\Big)\psi^{(6)}(2,3,1)\Big]\Big\}\ .
\end{align}
The sum of the $u$ and $d$ contributions in the $l_z=2$ component to the total orbital angular momentum is
\begin{align}
\ell_z^{l_z=2}=&\int\left[\ud x\right]_3\left[\ud^2\uvec k\right]_3\Big\{\psi^{(6)}(1,2,3)\nonumber\\
&\times\Big[2\uvec k_3^{\, 2}\uvec k_1^{\, 2}\Big(2\psi^{(6)}(1,2,3)+\psi^{(6)}(3,2,1)\Big)-2\uvec k_3^{\, 2}\uvec k_1\cdot\uvec k_2\Big(\psi^{(6)}(3,1,2)-\psi^{(6)}(2,1,3)\Big)\nonumber\\
&-2\uvec k_1^{\, 2}\uvec k_2\cdot\uvec k_3\Big(2\psi^{(6)}(1,3,2)+\psi^{(6)}(2,3,1)\Big)\Big]\Big\}\nonumber\\
=&\,\,2\,\rho_{l_z=2}.
\end{align}

\section{Results in light-cone quark models}
\label{sec:5}

The method developed in the previous sections is applied here within two light-cone quark models, a light-cone constituent quark model (LCCQM)~\cite{Boffi:2007yc,Pasquini:2005dk,Pasquini:2008ax,Pasquini:2010af,Pasquini:2007iz} and the light-cone version of the chiral quark-soliton model (LC$\chi$QSM) restricted to the three-quark sector \cite{Petrov:2002jr,Diakonov:2005ib,Lorce:2006nq}. These two models were recently applied to describe the valence-quark structure of the nucleon as observed in parton distribution functions, like generalized parton distributions, transverse-momentum dependent parton distributions, and form factors of the nucleon, giving a typical accuracy of about $30\%$ in comparison with available data in the valence region.

In the LCCQM the nucleon state is described by a LCWF in the basis of three free on-shell valence quarks. The three-quark state is however not on-shell $M_N\neq \mathcal{M}_0=\sum_i\omega_i$ where $\omega_i$ is the energy of free quark $i$ and $M_N$ is the physical mass of the nucleon bound state. The nucleon wave function is assumed to be a simple analytic function depending on three free parameters (including the quark mass) which are fitted to reproduce at best some experimental observables, like \emph{e.g.} the anomalous magnetic moments of the proton and neutron and the axial charge.

In the LC$\chi$QSM quarks are not free, but bound by a relativistic chiral mean field (semi-classical approximation). This chiral mean field creates a discrete level in the one-quark spectrum and distorts at the same time the Dirac sea. It has been shown that the distortion can be represented by additional quark-antiquark pairs in the baryon~\cite{Petrov:2002jr}. Even though the $\chi$QSM naturally incorporates higher Fock states, we restrict the present study to the 3Q sector.

Despite the apparent differences between the LCCQM and the $\chi$QSM, it turns out that the corresponding LCWFs have a very similar structure (for further details, we refer to~\cite{Lorce:2011dv,Lorce:2011zt}). The corresponding predictions from the LCCQM and the LC$\chi$QSM for $u$-, $d$- and total ($u+d$) quark contributions to the OAM are reported in Table~\ref{OAMtable}.
\begin{table}[t]
\begin{center}
\caption{\footnotesize{The results for the quark orbital angular momentum from the LCCQM and the LC$\chi$QSM for $u$-, $d$- and total ($u+d$) quark contributions.}}\label{OAMtable}
\begin{tabular}{@{\quad}c@{\quad}c@{\quad}|@{\quad}c@{\quad}c@{\quad}c@{\quad}|@{\quad}c@{\quad}c@{\quad}c@{\quad}}
\hline\hline
\multicolumn{2}{@{\quad}c@{\quad}|@{\quad}}{Model}&\multicolumn{3}{c@{\quad}|@{\quad}}{LCCQM}&\multicolumn{3}{c@{\quad}}{LC$\chi$QSM}\\
\multicolumn{2}{@{\quad}c@{\quad}|@{\quad}}{$q$}&$u$&$d$&Total&$u$&$d$&Total\\
\hline
$\ell^q_z$&Eq.~\eqref{lzform}&$0.131$&$-0.005$&$0.126$&$0.073$&$-0.004$&$0.069$\\
\hline\hline
\end{tabular}
\end{center}
\end{table}
We note that there is more net quark OAM in the LCCQM ($\sum_q\ell^q_z=0.126$) than in the LC$\chi$QSM ($\sum_q\ell^q_z=0.069$). For the individual quark contributions, both the LCCQM and the LC$\chi$QSM predict that $\ell^q_{z}$ are positive for $u$ quarks and negative for $d$ quarks, with the $u$-quark contribution larger than the $d$-quark contribution in absolute value. These results refer to the low hadronic scales of the models. Before making a meaningful comparison with results from experiments or lattice calculations which are usually obtained at higher scales, a proper treatment of the effects due to QCD evolution is essential~\cite{Altenbuchinger:2010sz}. However, this is beyond the scope of the present paper.

\begin{figure}[ht]
\begin{center}
\epsfig{file=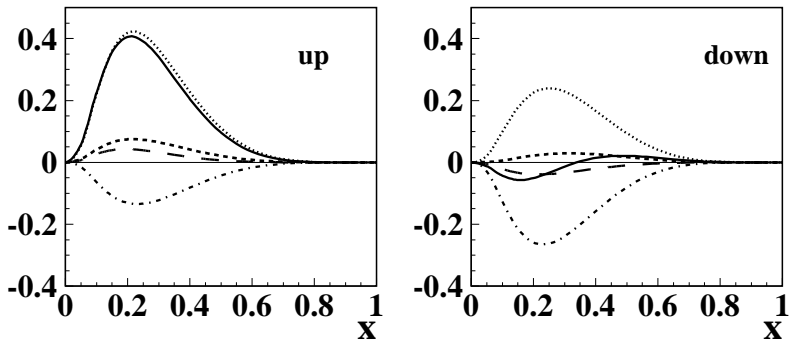, width=0.7\columnwidth}
\end{center}
\caption{Results for the distribution in $x$ of the OAM $\ell_z^q$ in the proton for $u$ (left) and $d$ (right) quark. The curves correspond to the contribution of the different partial waves: long-dashed curves for the light-cone amplitude with $l_z=0$, dotted curves for the light-cone amplitude with $l_z=1$, dashed-dotted curves for the light-cone amplitude with $l_z=-1$, and short-dashed curves for light-cone amplitude with $l_z=2$. The solid curves show the total results, sum of all the partial-wave contributions.}
\label{fig1}
\end{figure}

The explicit calculation within the LCCQM of the wave-function amplitudes in Eqs.~\eqref{lca1}-\eqref{lca4} can be found in Ref.~\cite{Pasquini:2008ax}. Using these results and the expressions in the appendix, we can also calculate the distribution in $x$ of the OAM $\ell_z^q$, separating the contribution from each partial wave. The corresponding results for $u$ and $d$ quarks are shown in Fig.~\ref{fig1}.
The $x$ dependence of the different partial-wave amplitudes is very similar for $u$ and $d$ quarks. However, the total contribution has a distinctive behavior for $u$ and $d$ quarks, coming from a quite different interplay between the different partial-waves. For the $u$ quarks, the dominant contribution comes from the $l_z=1$ amplitude (dotted curve), with positive sign, while the positive contributions coming from the $l_z=0$ (long-dashed curve) and $l_z=2$ (short-dahsed curve) amplitudes are largely compensated by the negative contribution coming from the $l_z=-1$ amplitude (dashed-dotted curve). For the $d$ quarks, the OAM arises from the competition between the positive $l_z=1$ and $l_z=2$ contributions, and the negative $l_z=-1$ and $l_z=0$ contributions. As a result, the OAM for 
$d$ quarks is much smaller than for $u$ quarks, and goes from negative to positive values at $x\approx 0.3$.

The integral over $x$ of the different distributions in Fig~\ref{fig1} gives the value for $u$- and $d$-quark OAM reported in Table~\ref{results}. In the last row we also show the results for the squared norm of the different partial-waves $\rho_{l_z}$, giving the probability to find the proton in a three-quark state with eigenvalue $l_z$ of total OAM, according to Eq.~\eqref{norm}.
\begin{table}[t]
\begin{center}
\caption{\footnotesize{Results from the LCCQM for the contribution of the different partial waves to the OAM $\ell^q_z$. The first and second rows show the values for the $u$ and $d$ quarks, respectively, while in the third row are reported the sum of the $u$- and $d$-quark contributions. In the last row are the results for the squared norm of the different partial-waves, giving the probability to find the proton in a three-quark state with eigenvalue $l_z$ of total OAM.}}
\label{results}
\begin{tabular}{@{\quad}c@{\quad}|@{\quad}c@{\quad\quad}c@{\quad\quad}c@{\quad\quad}c@{\quad}|@{\quad}c@{\quad}}
\hline\hline
&$l_z=0$&$l_z=1$&$l_z=-1$&$l_z=2$&Total\\
\hline
$\ell^u_z$&$0.013$&$0.139$&$-0.046$&$0.025$&$0.131$\\
$\ell^d_z$&$-0.013$&$0.087$&$-0.090$&$0.011$&$-0.005$\\
\hline
$\ell_z$&$0$&$0.226$&$-0.136$&$0.036$&$0.126$\\
\hline
$\rho_{l_z}$&$0.620$&$0.226$&$0.136$&$0.018$&$1$\\
\hline\hline
\end{tabular}
\end{center}
\end{table}

\begin{figure}[htb]
\includegraphics[width=0.45\textwidth]{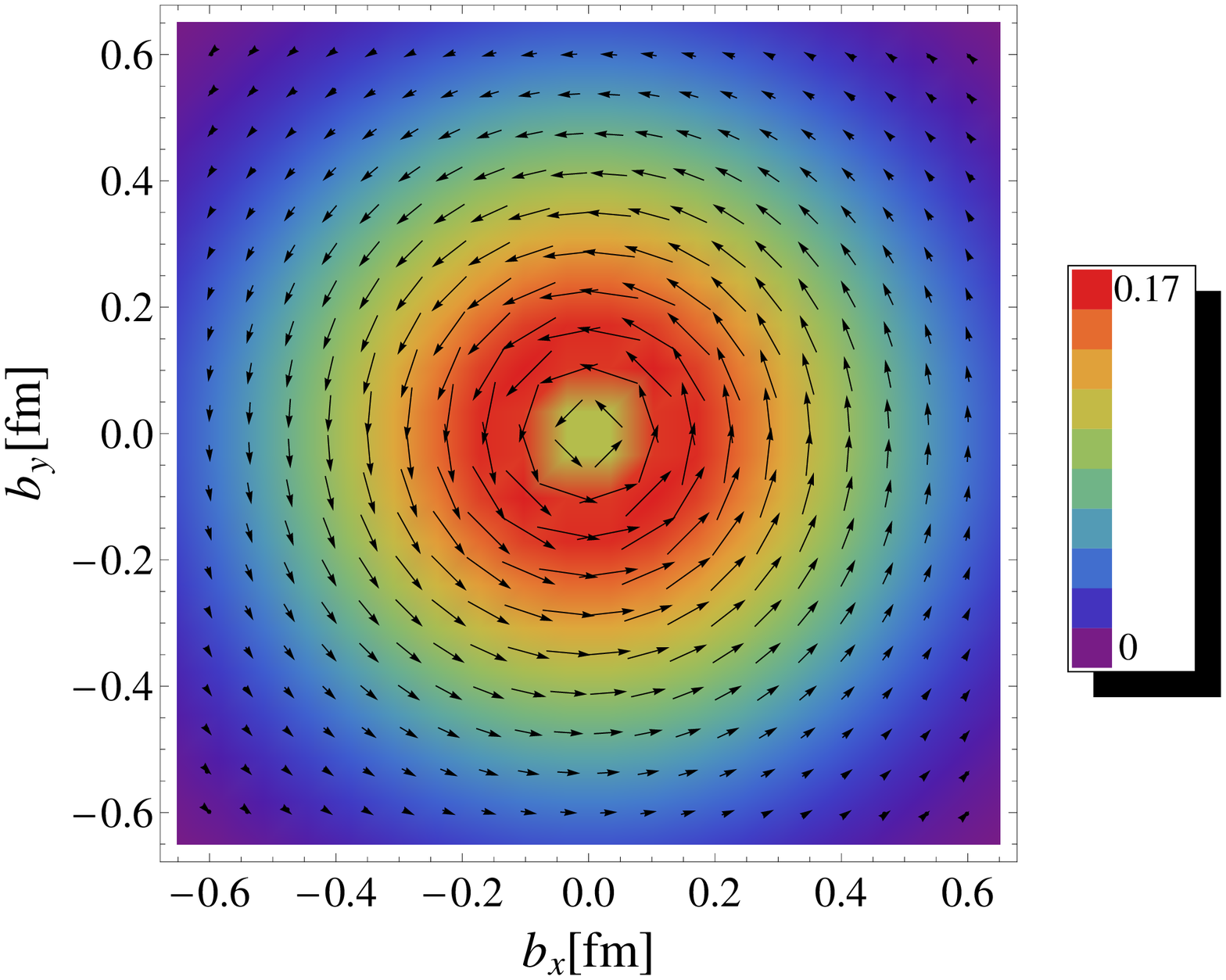}
\includegraphics[width=0.45\textwidth]{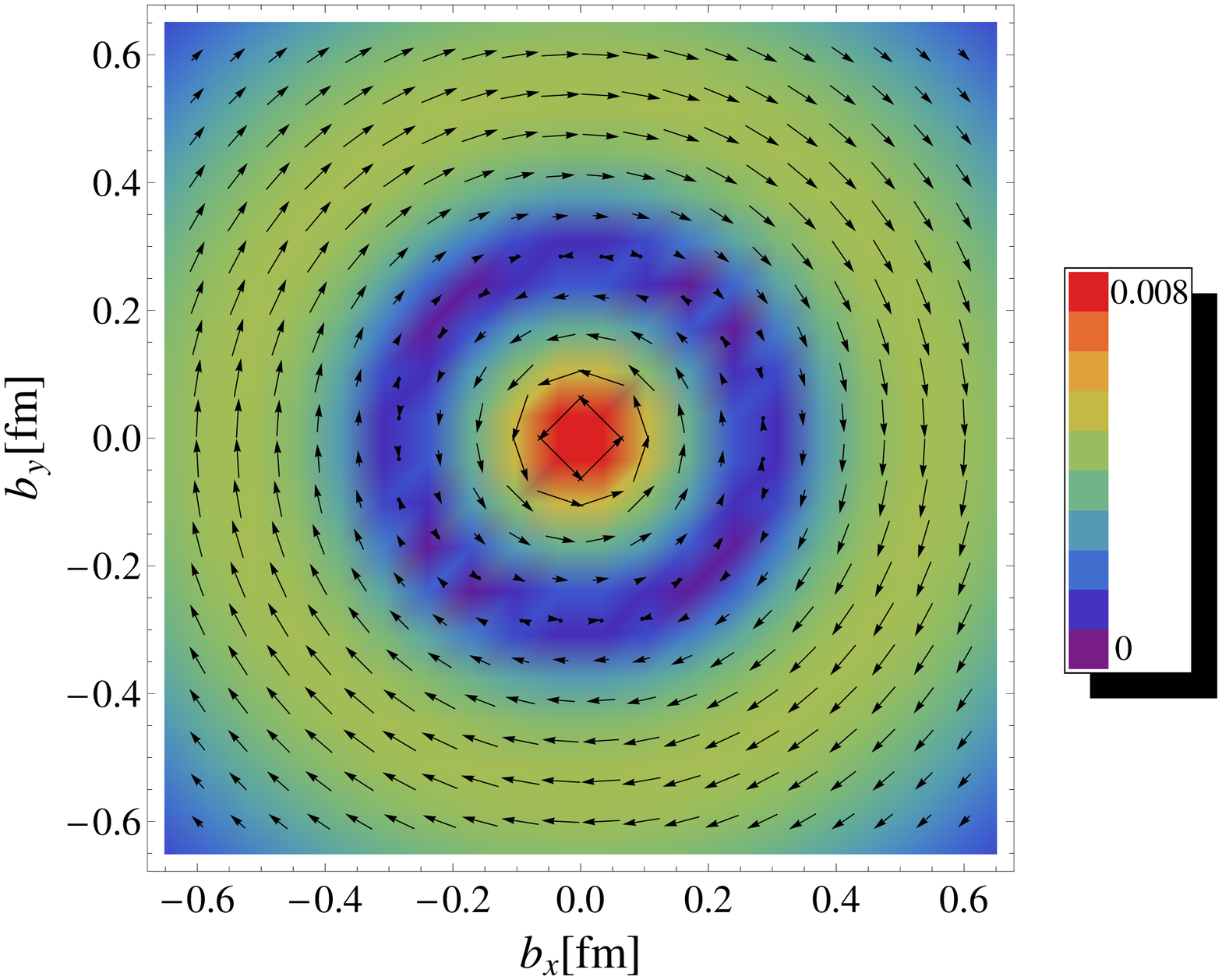}
\caption{Distributions in impact parameter space of the mean transverse momentum of unpolarized quarks in a longitudinally polarized nucleon. The nucleon polarization is pointing out of the plane, while the arrows show the size and direction of the mean transverse momentum of the quarks. The left (right) panel shows the results for $u$ ($d$) quarks.}
\label{run1} \vspace{-0mm}
\end{figure}

It is interesting to rewrite the expression \eqref{OAMform} for the OAM as
\be
\ell_z^q=\int\ud^2\uvec b\left(\uvec b\times\langle\uvec k\rangle^q\right)_z,
\ee
where $\langle\uvec k\rangle^q$ is the distribution in impact-parameter space of the quark mean transverse momentum
\be
\langle\uvec k\rangle^q(\uvec b)=\int\ud x\,\ud^2\uvec k\,\uvec k\,\rho^{[\gamma^+]q}_{++}(\uvec b,\uvec k,x).
\ee
This distribution is shown in Fig.~\ref{run1} for both $u$ and $d$ quarks. First of all, it appears that the mean transverse momentum $\langle\uvec k\rangle^q$ is always orthogonal to the impact-parameter vector $\uvec b$. This is not surprising since a nonvanishing radial component of the mean transverse momentum would indicate that the proton size and/or shape are changing. The figure also clearly shows that $u$ quarks tend to orbit anticlockwise inside the nucleon, corresponding to $\ell^u_z>0$ since the proton is represented with its spin pointing out of the figure. For the $d$ quarks, we see two regions. In the central region of the nucleon, $|\uvec b|<0.3$ fm, the $d$ quarks tend to orbit anticlockwise like the $u$ quarks. In the peripheral region, $|\uvec b|>0.3$ fm, the $d$ quarks tend to orbit clockwise. All this is consistent with the three-dimensional picture provided by the generalized parton distributions which indicates that the central region is dominated by the large $x$ values, while the peripheral region is dominated by the low $x$ values. The approximate cancellation between the central (large $x$) and peripheral (small $x$) contributions leads then to a very small value for the $d$-quark OAM.

\begin{figure}[htb]
\includegraphics[width=0.7\textwidth]{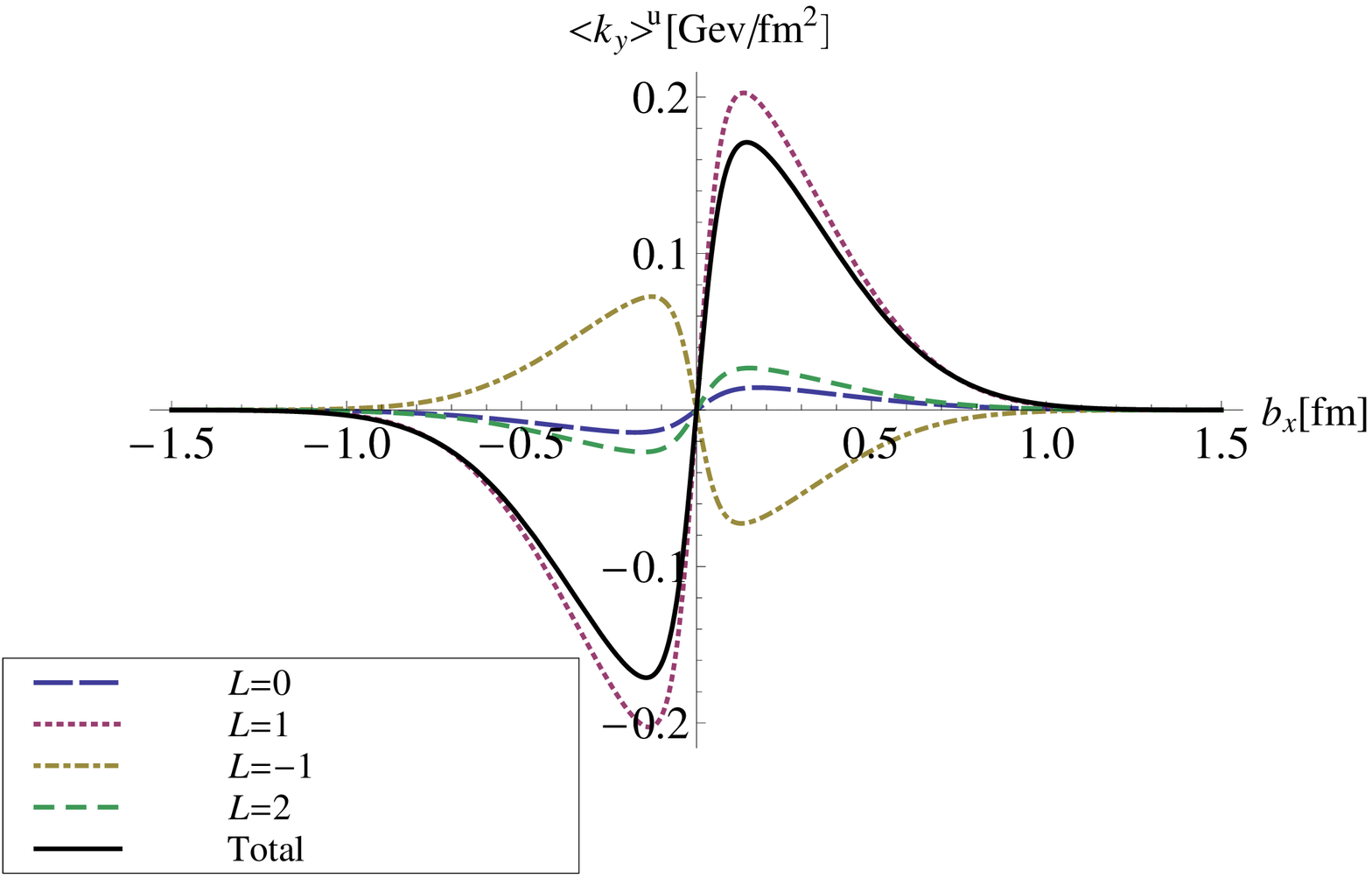}
\caption{Results for the different partial-wave contributions to the $u$ quark mean transverse momentum $\langle\uvec k\rangle^u=\langle k_y\rangle^u\,\uvec e_y$ as a function of $\uvec b=b_x\,\uvec e_x$.}
\label{ukr} \vspace{-0mm}
\end{figure}

\begin{figure}[htb]
\includegraphics[width=0.7\textwidth]{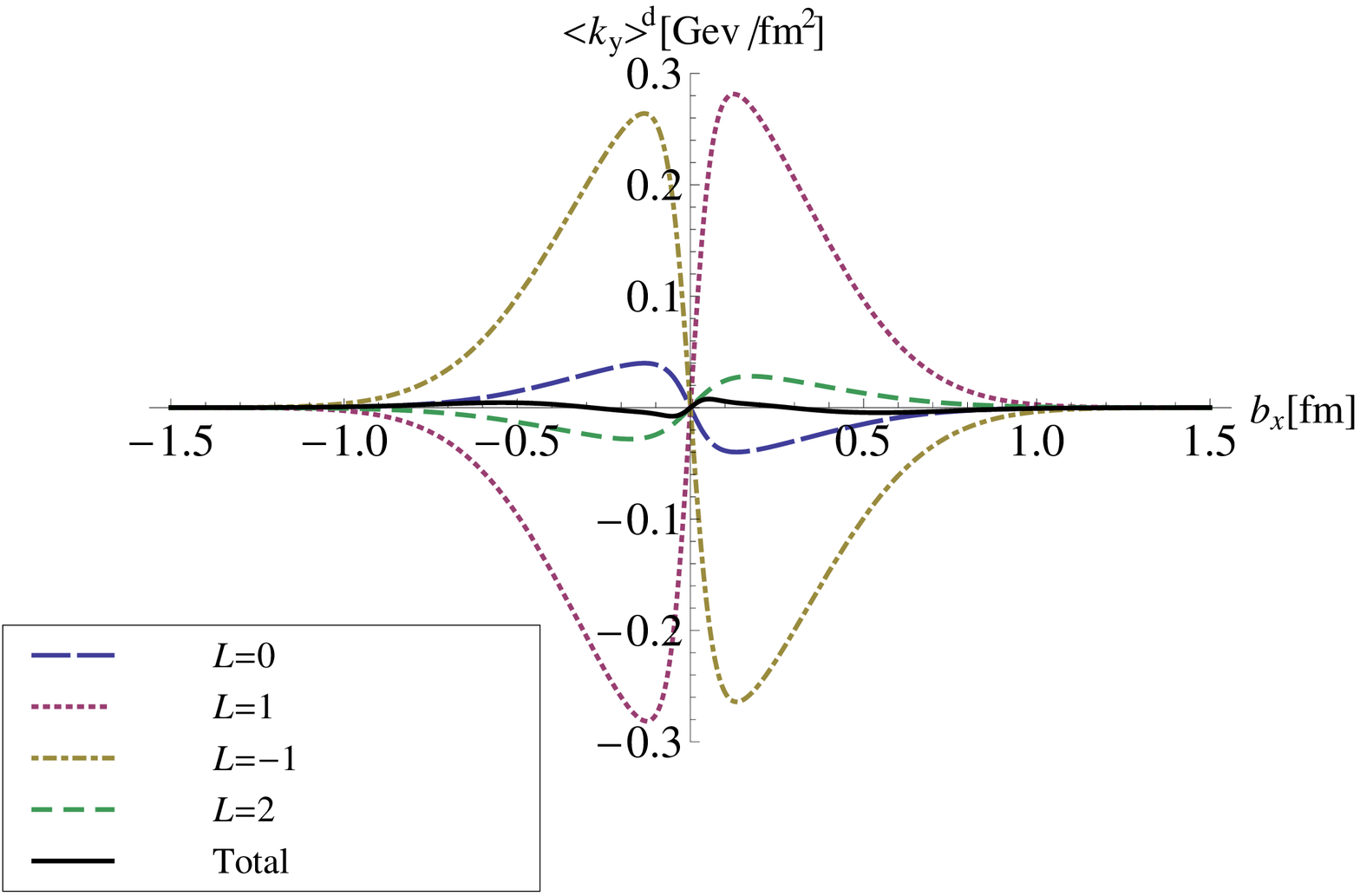}
\caption{Results for the different partial-wave contributions to the $d$ quark mean transverse momentum $\langle\uvec k\rangle^d=\langle k_y\rangle^d\,\uvec e_y$ as a function of $\uvec b=b_x\,\uvec e_x$.}
\label{dkr} \vspace{-0mm}
\end{figure}

Figs.~\ref{ukr} and \ref{dkr} show the different partial-wave contributions to the $u$ and $d$ quark mean transverse momentum $\langle\uvec k\rangle^q$ as a function of $\uvec b$. Due to the axial symmetry of the system, it is sufficient to plot $\langle\uvec k \rangle^q=\langle k_y\rangle^q\,\uvec e_y$ as a function of $\uvec b=b_x\,\uvec e_x$. Similarly to Fig.~\ref{fig1}, for the $u$ quarks, the dominant contribution comes from the $l_z=1$ amplitude (dotted curve), with positive sign, while the positive contributions coming from the $l_z=0$ (long-dashed curve) and $l_z=2$ (short-dahsed curve) amplitudes are largely compensated by the negative contribution coming from the $l_z=-1$ amplitude (dashed-dotted curve). For the $d$ quarks, the OAM arises from the competition between the positive $l_z=1$ and $l_z=2$ contributions, and the negative $l_z=-1$ and $l_z=0$ contributions, with a delicate balance between the different partial-wave contributions. As a result the total OAM takes small positive values at small $|\uvec b|$, becomes slightly negative at $|\uvec b|\approx 0.3 $ fm, and vanishes at the periphery.

\section{Conclusions}
\label{sec:6}

In summary, we derived the relation between the quark OAM and the Wigner distribution for unpolarized quark in a longitudinally polarized nucleon. This relation is exact as long as we neglect the contribution of gauge fields, and provides an intuitive and simple representation of the quark OAM which resembles the classical formula given by the phase-space average of the orbital angular momentum weighted by the density operator. We compare this derivation with the LCWF representation of the OAM. The advantage in using LCWFs is that they are eigenstates of the total OAM for each $N$-parton configuration in the nucleon Fock space.
As a consequence, the total OAM can be simply calculated from the sum of  squared  LCWFs multiplied by the corresponding 
eigenvalues of the OAM operator.
In the three-quark sector,  we further decomposed the nucleon state in different partial-wave amplitudes, calculating the corresponding contributions to the 
quark OAM.

These two representations of the OAM are equivalent, and allow to visualize
complementary aspects of the orbital motion of the quarks inside the nucleon.
As examples, we adopted two different light-cone quark models, and
 discussed the corresponding results for the 
distribution in $x$ of the
OAM, as obtained from the LCWF overlap representation, as well as the distribution of the mean transverse momentum in the impact-parameter space, as obtained from the Wigner distributions.

\vspace{3 cm}

\begin{acknowledgements}
This work was supported in part by the U.S. Department of Energy under contracts DE-AC02-05CH11231, 
and by the Italian MIUR through the PRIN 2008EKLACK ``Structure of the nucleon: transverse momentum, transverse spin and orbital angular momentum''. 
\end{acknowledgements}

\appendix

\section{Partial-wave decomposition of the distributions in $x$ of the OAM}
\label{appendix}

In this appendix we summarize the results for the partial-wave contributions to the distribution in $x$ of the OAM. In particular, we separately list the results for $\ell_z^{q_i,l_z}$, corresponding to the contribution of the $i$th quark with flavor $q$ in the Fock state component with total OAM $l_z$.

For the $l_z=0$ component, we find:
\\
${\bullet}$ for the $u$ quark
\begin{align}
&\ell_z^{u_1,l_z=0}(x)=\int\left[\ud x\right]_3\left[\ud^2\uvec k\right]_3\delta(x-x_1)\Big\{-\psi^{(1)}(1,2,3)\nonumber\\
&\times\Big[2\Big((1-x_1)\uvec k_1\cdot\uvec k_2+x_2\uvec k_1^{\, 2}\Big)\psi^{(2)}(1,2,3)+\Big(x_2\uvec k_1\cdot\uvec k_3-x_3\uvec k_1\cdot\uvec k_2\Big)\psi^{(2)}(3,2,1)\Big]\nonumber\\
&-(\uvec k_1\times\uvec k_2)\Big[\psi^{(1)}(1,2,3)\Big(\uvec k_1\times\widetilde{\uvec\nabla}_1\Big)\Big(-2\psi^{(2)}(1,2,3)+\psi^{(2)}(3,2,1)\Big)\nonumber\\
&+\psi^{(2)}(1,2,3)\Big(\uvec k_1\times\widetilde{\uvec\nabla}_1\Big)\Big(2\psi^{(1)}(1,2,3)+\psi^{(1)}(3,2,1)\Big)\Big]\Big\}\, ,
\end{align}
\begin{align}
&\ell_z^{u_2,l_z=0}(x)=\int\left[\ud x\right]_3\left[\ud^2\uvec k\right]_3\delta(x-x_2)\Big\{\psi^{(1)}(1,2,3)\Big[(1-x_2)\uvec k_1\cdot\uvec k_2+x_1\uvec k_2^{\, 2}\Big]\psi^{(2)}(1,2,3)\nonumber\\
&+(\uvec k_1\times\uvec k_2)\Big[\psi^{(1)}(1,2,3)\Big(\uvec k_2\times\widetilde{\uvec\nabla}_2\Big)\psi^{(2)}(1,2,3)-\psi^{(2)}(1,2,3)\Big(\uvec k_2\times\widetilde{\uvec\nabla}_2\Big)\psi^{(1)}(1,2,3)\Big]\Big\}\ ,
\end{align}
\begin{align}
&\ell_z^{u_3,l_z=0}(x)=\int\left[\ud x\right]_3\left[\ud^2\uvec k\right]_3\delta(x-x_3)\Big\{-\psi^{(1)}(1,2,3)\nonumber\\
&\times\Big[\Big(x_2\uvec k_1\cdot\uvec k_3-x_1\uvec k_2\cdot\uvec k_3\Big)\psi^{(2)}(1,2,3)+\Big((1-x_3)\uvec k_2\cdot\uvec k_3+x_2\uvec k_3^{\, 2}\Big)\psi^{(2)}(3,2,1)\Big]\nonumber\\
&-(\uvec k_1\times\uvec k_2)\Big[\psi^{(1)}(1,2,3)\Big(\uvec k_3\times\widetilde{\uvec\nabla}_3\Big)\Big(-\psi^{(2)}(1,2,3)+\psi^{(2)}(3,2,1)\Big)\nonumber\\
&+\psi^{(2)}(1,2,3)\Big(\uvec k_3\times\widetilde{\uvec\nabla}_3\Big)\Big(\psi^{(1)}(1,2,3)+\psi^{(1)}(3,2,1)\Big)\Big]\Big\};
\end{align}
\\
${\bullet}$ for the $d$ quark
\begin{align}
&\ell_z^{d_2,l_z=0}(x)=\int\left[\ud x\right]_3\left[\ud^2\uvec k\right]_3\delta(x-x_2)\Big\{\psi^{(1)}(1,2,3)\nonumber\\
&\times\Big[\Big((1-x_2)\uvec k_1\cdot\uvec k_2+x_1\uvec k_2^{\, 2}\Big)\psi^{(2)}(1,2,3)+\Big(x_3\uvec k_2^{\, 2}+(1-x_2)\uvec k_2\cdot\uvec k_3\Big)\psi^{(2)}(3,2,1)\Big]\nonumber\\
&-(\uvec k_1\times\uvec k_2)\Big[\psi^{(1)}(1,2,3)\Big(\uvec k_2\times\widetilde{\uvec\nabla}_2\Big)\Big(-\psi^{(2)}(1,2,3)+\psi^{(2)}(3,2,1)\Big)\nonumber\\
&+\psi^{(2)}(1,2,3)\Big(\uvec k_2\times\widetilde{\uvec\nabla}_2\Big)\Big(\psi^{(1)}(1,2,3)+\psi^{(1)}(3,2,1)\Big)\Big]\Big\},
\end{align}
\begin{align}
&\ell_z^{d_3,l_z=0}(x)=\int\left[\ud x\right]_3\left[\ud^2\uvec k\right]_3\delta(x-x_3)\Big\{-(x_2\uvec k_1\cdot\uvec k_3-x_1\uvec k_2\cdot\uvec k_3)\psi^{(1)}(1,2,3)\psi^{(2)}(1,2,3)\nonumber\\
&+(\uvec k_1\times\uvec k_2)\Big[\psi^{(1)}(1,2,3)\Big(\uvec k_3\times\widetilde{\uvec\nabla}_3\Big)\psi^{(2)}(1,2,3)-\psi^{(2)}(1,2,3)\Big(\uvec k_3\times\widetilde{\uvec\nabla}_3\Big)\psi^{(1)}(1,2,3)\Big]\Big\}\ .
\end{align}

For the $l_z=1$ component, we find:
\\
${\bullet}$ for the $u$ quark
\begin{align}
&\ell_z^{u_1,l_z=1}(x)=\int\left[\ud x\right]_3\left[\ud^2\uvec k\right]_3\delta(x-x_1)\nonumber\\
&\Big\{\Big[\uvec k_1^{\, 2}\psi^{(3)}(1,2,3)) +\uvec k_1\cdot \uvec k_2\psi^{(4)}(1,2,3)\Big]\Big[(1-x_1)\psi^{(3)}(1,2,3)-x_2\psi^{(4)}(1,2,3)\Big]\nonumber\\
&+(\uvec k_1\times\uvec k_2)\Big[\psi^{(3)}(1,2,3)\Big(\uvec k_1\times\widetilde{\uvec\nabla}_1\Big)\vec\psi^{(4)}(1,2,3)-\psi^{(4)}(1,2,3)\Big(\uvec k_1\times\widetilde{\uvec\nabla}_1\Big)\vec\psi^{(3)}(1,2,3)\Big]\Big\}\ ,
\end{align}
\begin{align}
&\ell_z^{u_2,l_z=1}(x)=\int\left[\ud x\right]_3\left[\ud^2\uvec k\right]_3\delta(x-x_2)\Big\{\Big[\uvec k_1\cdot\uvec k_2\psi^{(3)}(1,2,3))+\uvec k_2^{\, 2}\psi^{(4)}(1,2,3)\Big]\nonumber\\
&\times\Big[-2x_1\psi^{(3)}(1,2,3)+2(1-x_2)\psi^{(4)}(1,2,3)-x_1\psi^{(3)}(1,3,2)-x_3\psi^{(4)}(1,3,2)\Big]\nonumber\\
&-(\uvec k_1\times\uvec k_2)\Big[\psi^{(3)}(1,2,3)\Big(\uvec k_2\times\widetilde{\uvec\nabla}_2\Big)\Big(-2\psi^{(4)}(1,2,3)+\psi^{(4)}(1,3,2)\Big)\nonumber\\
&+\psi^{(4)}(1,2,3)\Big(\uvec k_2\times\widetilde{\uvec\nabla}_2\Big)\Big(2\psi^{(3)}(1,2,3)+\psi^{(3)}(1,3,2)\Big)\nonumber\\
&-\psi^{(4)}(1,2,3)\Big(\uvec k_2\times\widetilde{\uvec\nabla}_2\Big)\psi^{(4)}(1,3,2)\Big]\Big\}\ ,
\end{align}
\begin{align}
&\ell_z^{u_3,l_z=1}(x)=\int\left[\ud x\right]_3\left[\ud^2\uvec k\right]_3\delta(x-x_3)\Big\{\Big[\uvec k_1\cdot\uvec k_3\psi^{(3)}(1,2,3))+\uvec k_2\cdot \uvec k_3\psi^{(4)}(1,2,3)\Big]\nonumber\\
&\times\Big[-x_1\psi^{(3)}(1,2,3)-x_2\psi^{(4)}(1,2,3)-x_1\psi^{(3)}(1,3,2)+(1-x_3)\psi^{(4)}(1,3,2)\Big]\nonumber\\
&-(\uvec k_1\times\uvec k_2)\Big[\psi^{(3)}(1,2,3)\Big(\uvec k_3\times\widetilde{\uvec\nabla}_3\Big)\Big(-\psi^{(4)}(1,2,3)+\psi^{(4)}(1,3,2)\Big)\nonumber\\
&+\psi^{(4)}(1,2,3)\Big(\uvec k_3\times\widetilde{\uvec\nabla}_3\Big)\Big(\psi^{(3)}(1,2,3)+\psi^{(3)}(1,3,2)\Big)\nonumber\\
&-\psi^{(4)}(1,2,3)\Big(\uvec k_3\times\widetilde{\uvec\nabla}_3\Big)\psi^{(4)}(1,3,2)\Big]\Big\}\ ;
\end{align}
\\
${\bullet}$ for the $d$ quark
\begin{align}
&\ell_z^{d_1,l_z=1}(x)=\int\left[\ud x\right]_3\left[\ud^2\uvec k\right]_3\delta(x-x_1)\Big\{\Big[\uvec k_1^{\, 2}\psi^{(3)}(1,2,3))+\uvec k_1\cdot \uvec k_2\psi^{(4)}(1,2,3)\Big]\nonumber\\
&\times\Big[(1-x_1)\psi^{(3)}(1,2,3)-x_2\psi^{(4)}(1,2,3)+(1-x_{1})\psi^{(3)}(1,3,2)-x_3\psi^{(4)}(1,3,2)\Big]\nonumber\\
&-(\uvec k_1\times\uvec k_2)\Big[\psi^{(3)}(1,2,3)\Big(\uvec k_1\times\widetilde{\uvec\nabla}_1\Big)\Big(-\psi^{(4)}(1,2,3)+\psi^{(4)}(1,3,2)\Big)\nonumber\\
&+\psi^{(4)}(1,2,3)\Big(\uvec k_1\times\widetilde{\uvec\nabla}_1\Big)\Big(\psi^{(3)}(1,2,3)+\psi^{(3)}(1,3,2)\Big)\nonumber\\
&-\psi^{(4)}(1,2,3)\Big(\uvec k_1\times\widetilde{\uvec\nabla}_1\Big)\psi^{(4)}(1,3,2))\Big]\Big\}\ ,
\end{align}
\begin{align}
&\ell_z^{d_3,l_z=1}(x)=\int\left[\ud x\right]_3\left[\ud^2\uvec k\right]_3\delta(x-x_3)\nonumber\\
&\Big\{\Big[\uvec k_1\cdot\uvec k_3\psi^{(3)}(1,2,3))+\uvec k_2\cdot\uvec k_3\psi^{(4)}(1,2,3)\Big]\Big[-x_1\psi^{(3)}(1,2,3)-x_2\psi^{(4)}(1,2,3)\Big]\nonumber\\
&+(\uvec k_1\times\uvec k_2)\Big[\psi^{(3)}(1,2,3)\Big(\uvec k_3\times\widetilde{\uvec\nabla}_3\Big)\psi^{(4)}(1,2,3))-\psi^{(4)}(1,2,3)\Big(\uvec k_3\times\widetilde{\uvec\nabla}_3\Big)\psi^{(3)}(1,2,3)\Big]\Big\}\ .
\end{align}

For $l_z=-1$ component, we find:
\\
${\bullet}$ for the $u$ quark
\begin{align}
&\ell_z^{u_1,l_z=-1}(x)=\int\left[\ud x\right]_3\left[\ud^2\uvec k\right]_3\delta(x-x_1)\Big\{\uvec k_1\cdot\uvec k_2\psi^{(5)}(1,2,3)\nonumber\\
&\times\Big[2x_2\psi^{(5)}(1,2,3)-(1-x_1)\psi^{(5)}(2,1,3)+x_2\psi^{(5)}(3,2,1) -2x_3\psi^{(5)}(1,3,2)\nonumber\\
&+(1-x_1)\psi^{(5)}(3,1,2)-x_3\psi^{(5)}(2,3,1)\Big]-(\uvec k_1\times\uvec k_2)\psi^{(5)}(1,2,3)\nonumber\\
&\times\Big(\uvec k_1\times\widetilde{\uvec\nabla}_1\Big)\Big[\psi^{(5)}(3,1,2)-\psi^{(5)}(2,1,3)-2\psi^{(5)}(1,3,2)+\psi^{(5)}(2,3,1)\Big]\Big\}\ ,
\end{align}
\begin{align}
&\ell_z^{u_2,l_z=-1}(x)=\int\left[\ud x\right]_3\left[\ud^2\uvec k\right]_3\delta(x-x_2)\Big\{\uvec k_2^{\; 2}\psi^{(5)}(1,2,3)\nonumber\\
&\times\Big[-(1-x_2)\psi^{(5)}(1,2,3)+x_1\psi^{(5)}(2,1,3)-x_3\psi^{(5)}(1,3,2) -x_3\psi^{(5)}(2,3,1)\Big]\nonumber\\
&+(\uvec k_1\times\uvec k_2)\psi^{(5)}(1,2,3)\Big(\uvec k_2\times\widetilde{\uvec\nabla}_2\Big)\Big[\psi^{(5)}(2,1,3)+\psi^{(5)}(1,3,2)+\psi^{(5)}(2,3,1)\Big]\Big\}\ ,
\end{align}
\begin{align}
&\ell_z^{u_3,l_z=-1}(x)=\int\left[\ud x\right]_3\left[\ud^2\uvec k\right]_3\delta(x-x_3)\Big\{\uvec k_2\cdot\uvec k_3\psi^{(5)}(1,2,3)\nonumber\\
&\Big[x_2\psi^{(5)}(1,2,3)+x_2\psi^{(5)}(3,2,1)+(1-x_3)\psi^{(5)}(1,3,2) -x_1\psi^{(5)}(3,1,2)\Big]\nonumber\\
&-(\uvec k_1\times\uvec k_2)\psi^{(5)}(1,2,3)\Big(\uvec k_3\times\widetilde{\uvec\nabla}_3\Big)\Big[\psi^{(5)}(3,1,2))-\psi^{(5)}(1,3,2)\Big]\Big\}\ ;
\end{align}
\\
${\bullet}$ for the $d$ quark
\begin{align}
&\ell_z^{d_2,l_z=-1}(x)=\int\left[\ud x\right]_3\left[\ud^2\uvec k\right]_3\delta(x-x_2)\Big\{-\uvec k_2^{\, 2}\psi^{(5)}(1,2,3)\nonumber\\
&\times\Big[(1-x_2)\psi^{(5)}(1,2,3)+(1-x_2)\psi^{(5)}(3,2,1)+x_3\psi^{(5)}(1,3,2) +x_1\psi^{(5)}(3,1,2)\Big]\nonumber\\
&-(\uvec k_1\times\uvec k_2)\psi^{(5)}(1,2,3)\Big(\uvec k_2\times\widetilde{\uvec\nabla}_2\Big)\Big[\psi^{(5)}(3,1,2)-\psi^{(5)}(1,3,2)\Big]\Big\}\ ,
\end{align}
\begin{align}
&\ell_z^{d_3,l_z=-1}(x)=\int\left[\ud x\right]_3\left[\ud^2\uvec k\right]_3\delta(x-x_3)\Big\{\uvec k_2\cdot\uvec k_3\psi^{(5)}(1,2,3)\nonumber\\
&\times\Big[x_1\psi^{(5)}(2,1,3)+x_2\psi^{(5)}(1,2,3)+(1-x_3)\psi^{(5)}(1,3,2) +(1-x_3)\psi^{(5)}(2,3,1)\Big]\nonumber\\
&+(\uvec k_1\times\uvec k_2)\psi^{(5)}(1,2,3)\Big(\uvec k_3\times\widetilde{\uvec\nabla}_3\Big)\Big[\psi^{(5)}(2,1,3)+\psi^{(5)}(1,3,2)+\psi^{(5)}(2,3,1)\Big]\Big\}\ .
\end{align}

For $l_z=2$ component, we find:
\\
${\bullet}$ for the $u$ quark
\begin{align}
&\ell_z^{u_1,l_z=2}(x)=\int\left[\ud x\right]_3\left[\ud^2\uvec k\right]_3\delta(x-x_1)\Big\{\uvec k_1^{\, 2}\psi^{(6)}(1,2,3)\nonumber\\
&\times\Big[(1-x_1)\uvec k_3^{\, 2}-x_3\uvec k_1\cdot\uvec k_3\Big]\Big[2\psi^{(6)}(1,2,3)+\psi^{(6)}(3,2,1)\Big]\nonumber\\
&+\Big[x_2\uvec k_3^{\, 2}+x_3\uvec k_2\cdot\uvec k_3\Big]\Big[\psi^{(6)}(3,1,2)-\psi^{(6)}(2,1,3)\Big]\nonumber\\
&-\Big[(1-x_1)\uvec k_2\cdot\uvec k_3-x_2\uvec k_1\cdot\uvec k_3\Big]\Big[2\psi^{(6)}(1,3,2)+\psi^{(6)}(2,3,1)\Big]\nonumber\\
&+(\uvec k_1\times\uvec k_2)\psi^{(6)}(1,2,3)\Big[\uvec k_3^{\, 2}\Big(\uvec k_1\times\widetilde{\uvec\nabla}_1\Big)\Big(\psi^{(6)}(2,1,3))-\psi^{(6)}(3,1,2)\Big)\nonumber\\
&+\uvec k_1^{\, 2}\Big(\uvec k_1\times\widetilde{\uvec\nabla}_1\Big)\Big(2\psi^{(6)}(1,3,2)+\psi^{(6)}(2,3,1)\Big)\Big]\Big\}\ ,
\end{align}
\begin{align}
&\ell_z^{u_2,l_z=2}(x)=\int\left[\ud x\right]_3\left[\ud^2\uvec k\right]_3\delta(x-x_2)\nonumber\\
&\Big\{-\Big[x_1\uvec k_3^{\, 2}\uvec k_1\cdot\uvec k_2+x_3\uvec k_1^{\, 2}\uvec k_2\cdot\uvec k_3\Big]\psi^{(6)}(1,2,3)\,\psi^{(6)}(1,2,3)\nonumber\\
&-\Big[(1-x_2)\uvec k_1^{\, 2}\uvec k_2\cdot\uvec k_3-x_1\Big(2\uvec k_1\cdot\uvec k_2\uvec k_2\cdot\uvec k_3-\uvec k_2^{\, 2}\uvec k_1\cdot\uvec k_3\Big)\Big]\nonumber\\
&\times\Big[\psi^{(6)}(1,2,3)\psi^{(6)}(1,3,2)+\psi^{(6)}(2,3,1)\Big(\psi^{(6)}(1,2,3)-\psi^{(6)}(3,2,1)\Big)\Big]\nonumber\\
&+(\uvec k_1\times\uvec k_2)\psi^{(6)}(1,2,3)\Big[\uvec k_3^{\, 2}\Big(\uvec k_2\times\widetilde{\uvec\nabla}_2\Big)\psi^{(6)}(2,1,3)\nonumber\\
&+\uvec k_1^{\, 2}\Big(\uvec k_2\times\widetilde{\uvec\nabla}_2\Big)\Big(\psi^{(6)}(1,3,2)+\psi^{(6)}(2,3,1)\Big)\Big]\Big\}\ ,
\end{align}
\begin{align}
&\ell_z^{u_3,l_z=2}(x)=\int\left[\ud x\right]_3\left[\ud^2\uvec k\right]_3\delta(x-x_3)\Big\{\psi^{(6)}(1,2,3)\uvec k_3^{\, 2}\nonumber\\
&\times\Big[\Big((1-x_3)\uvec k_1^{\, 2}-x_1\uvec k_1\cdot\uvec k_3\Big)\Big(\psi^{(6)}(3,2,1)+\psi^{(6)}(1,2,3)\Big)+\Big(x_1\uvec k_1\cdot\uvec k_2+x_2\uvec k_1^{\, 2}\Big)\psi^{(6)}(1,3,2)\nonumber\\
&-\Big((1-x_3)\uvec k_1\cdot\uvec k_2-x_2\uvec k_1\cdot\uvec k_3\Big)\psi^{(6)}(3,1,2)\Big]\nonumber\\
&-(\uvec k_1\times\uvec k_2)\psi^{(6)}(1,2,3)\Big[\uvec k_3^{\, 2}\Big(\uvec k_3\times\widetilde{\uvec\nabla}_3\Big)\psi^{(6)}(3,1,2)-\uvec k_1^{\, 2}\Big(\uvec k_3\times\widetilde{\uvec\nabla}_3\Big)\psi^{(6)}(1,3,2)\Big]\Big\}\ ;
\end{align}
\\
${\bullet}$ for the $d$ quark
\begin{align}
&\ell_z^{d_2,l_z=2}(x)=\int\left[\ud x\right]_3\left[\ud^2\uvec k\right]_3\delta(x-x_2)\nonumber\\
&\times\Big\{-\Big[x_1\uvec k_3^{\, 2}\uvec k_1\cdot\uvec k_2+x_3\uvec k_1^{\, 2}\uvec k_2\cdot\uvec k_3\Big]\psi^{(6)}(1,2,3)\Big[\psi^{(6)}(1,2,3)+\psi^{(6)}(3,2,1)\Big]\nonumber\\
&-\Big[(1-x_2)\uvec k_1^{\, 2}\uvec k_2\cdot\uvec k_3-x_1(2\uvec k_1\cdot\uvec k_2\uvec k_2\cdot\uvec k_3-\uvec k_1\cdot\uvec k_3\uvec k_2^{\, 2})\Big]\nonumber\\
&\times\psi^{(6)}(1,3,2)\Big[\psi^{(6)}(1,2,3)+\psi^{(6)}(3,2,1)\Big]\nonumber\\
&+(\uvec k_1\times\uvec k_2)\psi^{(6)}(1,2,3)\Big[\uvec k_1^{\, 2}\Big(\uvec k_2\times\widetilde{\uvec\nabla}_2\Big)\psi^{(6)}(1,3,2)-\uvec k_3^{\, 2}\Big(\uvec k_2\times\widetilde{\uvec\nabla}_2\Big)\psi^{(6)}(3,1,2)\Big]\Big\}\ ,
\end{align}
\begin{align}
&\ell_z^{d_3,l_z=2}(x)=\int\left[\ud x\right]_3\left[\ud^2\uvec k\right]_3\delta(x-x_3)\Big\{\uvec k_3^{\, 2}\psi^{(6)}(1,2,3)\nonumber\\
&\times\Big[\Big((1-x_3)\uvec k_1\cdot\uvec k_2-x_2\uvec k_1\cdot\uvec k_3\Big)\psi^{(6)}(2,1,3)-\Big(x_1\uvec k_1\cdot\uvec k_3-(1-x_3)\uvec k_1^{\, 2}\Big)\psi^{(6)}(1,2,3)\nonumber\\
&+\Big(x_1\uvec k_1\cdot\uvec k_2+x_2\uvec k_1^{\, 2}\Big)\Big(\psi^{(6)}(1,3,2)+\psi^{(6)}(2,3,1)\Big)\Big]+(\uvec k_1\times\uvec k_2)\psi^{(6)}(1,2,3)\nonumber\\
&\times\Big[\uvec k_1^{\, 2}\Big(\uvec k_3\times\widetilde{\uvec\nabla}_3\Big)\Big(\psi^{(6)}(1,3,2)+\psi^{(6)}(2,3,1)\Big)+\uvec k_3^{\, 2}\Big(\uvec k_3\times\widetilde{\uvec\nabla}_3\Big)\psi^{(6)}(2,1,3)\Big]\Big\}\ .
\end{align}

\end{document}